\documentclass[10pt,onecolumn]{article}
\usepackage{ISG_style_math}
\usepackage[utf8]{inputenc} 
\usepackage[T1]{fontenc}    
\usepackage{hyperref}       
\usepackage{url}            
\usepackage{booktabs}       
\usepackage{amsfonts}       
\usepackage{nicefrac}       
\usepackage{microtype}      

\usepackage{times}
\usepackage{graphicx}
\usepackage{subcaption}
\usepackage{xcolor}
\usepackage{amsmath,amssymb}
\usepackage{bbm}
\usepackage{dsfont}
\usepackage{cite}
\usepackage{flushend}
\usepackage{float}

\usepackage{algorithm}
\usepackage{algpseudocode}
\usepackage{mathtools}
\usepackage{appendix}

\DeclareMathOperator\supp{supp}

\DeclareMathOperator*{\argmin}{argmin}

\def \P {{\sf P }} 
\def \Q {{\sf Q }} 
\def \C {{\rm C }} 
\def \mP {{\mathbb{P}}}

\newcommand\mcGo{\mathcal{G}^{(1)}}
\newcommand\mcGt{\mathcal{G}^{(2)}}
\newcommand\Bo{B^{(1)}}
\newcommand\Bt{B^{(2)}}
\newcommand\eo{\epsilon^{(1)}}
\newcommand\et{\epsilon^{(2)}}
\newcommand\Thetao{\Theta^{(1)}}
\newcommand\Thetat{\Theta^{(2)}}

\newcommand\Sigmao{\Sigma^{(1)}}
\newcommand\Sigmat{\Sigma^{(2)}}
\newcommand\sigmao{\sigma^{(1)}}
\newcommand\sigmat{\sigma^{(2)}}

\newcommand\Algphase[1]{%
\vspace*{-.5\baselineskip}\Statex\hspace*{\dimexpr-\algorithmicindent-2pt\relax}\rule{\textwidth}{0.4pt}%
\Statex\hspace*{-\algorithmicindent}\textbf{#1}%
\vspace*{-.7\baselineskip}\Statex\hspace*{\dimexpr-\algorithmicindent-2pt\relax}\rule{\textwidth}{0.4pt}%
}

\newcommand\blfootnote[1]{%
  \begingroup
  \renewcommand\thefootnote{}\footnote{#1}%
  \addtocounter{footnote}{-1}%
  \endgroup
}

\title{Scalable Intervention Target Estimation in Linear Models}
\date{}
\author{Burak Var\i c\i \thanks{Rensselaer Polytechnic Institute} \\ \and Karthikeyan Shanmugam \thanks{IBM Research AI} \\ \and Prasanna Sattigeri \footnotemark[2] \\  \and Ali Tajer \footnotemark[1]}

\begin{document}

\maketitle
\blfootnote{To appear at 35th Conference on Neural Information Processing Systems (NeurIPS 2021)}
\begin{abstract}
This paper considers the problem of estimating the unknown intervention targets in a causal directed acyclic graph from observational and interventional data. The focus is on soft interventions in linear structural equation models (SEMs). Current approaches to causal structure learning either work with known intervention targets or use hypothesis testing to discover the unknown intervention targets even for linear SEMs. This severely limits their scalability and sample complexity. This paper proposes a scalable and efficient algorithm that consistently identifies all intervention targets. The pivotal idea is to estimate the intervention sites from the difference between the precision matrices associated with the observational and interventional datasets. It involves repeatedly estimating such sites in different subsets of variables. The proposed algorithm can be used to also update a given observational Markov equivalence class into the interventional Markov equivalence class. Consistency, Markov equivalency, and sample complexity are established analytically. Finally, simulation results on both real and synthetic data demonstrate the gains of the proposed approach for scalable causal structure recovery. Implementation of the algorithm and the code to reproduce the simulation results are available at \url{https://github.com/bvarici/intervention-estimation}. 

\end{abstract}

\section{Introduction}\label{sec:introduction}
Directed acyclic graphs (DAG) are commonly used for encoding the cause-effect relationships among random variables. Extensive research has been dedicated to learning the structure of DAGs from their associated observational data. Structure learning from the observational data relies on uncovering conditional independence (CI) among the random variables. Since structurally distinct DAGs can encode the same set of CI relations, a DAG is identifiable only up to its Markov equivalence class (MEC) from the observational data. Subsequently, \emph{interventional} data can be used to further refine the MEC obtained from the observational data and learn specific causal effects.

This paper is motivated by addressing two significant independent challenges in causal discovery. First, most of the existing approaches for learning with interventional datasets require the \emph{intervention target set} to be known, which can be a strong assumption. For instance, gene-editing technologies are known to perform cleavage at off-target genome sites \cite{fu2013high}. Therefore, identifying the \emph{intervened} nodes alone is a critical problem in structure learning, and despite its significance, it remains uninvestigated. Secondly, besides learning the structures of single DAGs, there exist application domains in which the goal is learning the structural changes between two related networks and their associated DAGs, or learning the sites of interventions. For instance, structural differences between the gene regulatory networks of different subtypes of cancers can help to identify the roles of specific genes \cite{ovariancancer}. In electroencephalography analysis, the objective is to detect different brain regions that have different interactions when the subject is performing various tasks \cite{sanei2013eeg}. These brain regions correspond to intervened nodes in a causal graph representation. Another application area is fault detection in large-scale Internet of things and cloud applications \cite{bogatinovski2021artificial}. Faulty nodes in the system can be considered as intervened nodes, and they are localized through intervention target estimation.

In practice, fixing the target variable at a specific value or removing its causal dependencies is often difficult, while disturbing the distribution of a target variable is easier \cite{eberhardt2007interventions}. The type of interventions that do not remove the causal effects are commonly observed in the real world. For example, elements of an advertising system can be modified without removing the causal effects \cite{bottou2013counterfactual}. In another example, consider molecular biology, in which the effects of infused chemicals to the cell are not set to specific values nor are they known precisely \cite{jaber2020causal,eaton2007}. Therefore, we consider a \emph{soft intervention} setting, in which we assume the conditional distributions of the target variables are changed, but no assumption is made on the causal effects. Finally, we assume that the topological ordering remains the same after the intervention procedure.

Under the soft intervention model, we propose an algorithm for estimating the intervention targets given the data from two linear SEMs associated with the observational and interventional data. Motivated by the fact that the difference of the precision matrices associated with these two models is sparse, we focus on estimating the sparse difference between precision matrices to avoid extensive conditional independence testing. This leads to a significant improvement in the computational complexity compared to those of the alternative methods. This facilitates scaling up to high-dimensional settings. Furthermore, we show that this algorithm can be used in conjunction with an observational DAG learning procedure to refine the MEC to \emph{interventional}-MEC ($\mcI$-MEC). Besides being consistent in the population setting, we provide the finite-sample guarantees for linear SEM with Gaussian noise when the soft interventional changes between the two models are sparse. Our main contributions are as follows:
\begin{itemize}
    \item We propose an algorithm that identifies intervention targets under the intervention-faithfulness assumption. We show that our algorithm identifies $\mcI$-MEC given the observational MEC.
    \item We provide the sample complexity of our algorithm under linear SEM with Gaussian noise.
    \item We perform experiments on both real biological and synthetic datasets to illustrate the ability to work in the high-dimensional settings and the gains compared to the relevant methods. 
\end{itemize}
\section{Related work} \label{sec:related_work}
Among the broad range of approaches to intervention recovery, there exist two methods that are closely related to the scope of this paper: (i) estimating the difference between two DAGs, and (ii) learning from a combination of observational and interventional datasets.

\textbf{Direct estimation of differences:} Direct estimation of differences in linear SEMs has been studied recently. The study in \cite{wang2018direct} proposes a PC-style algorithm for learning changes in the edge weights by testing invariances of regression coefficients and noise variances. Even though the differences can be sparse, individual models can be dense, and estimating these variables through regression can be inaccurate. Furthermore, the number of hypothesis tests is exponential in the number of nodes that are affected by the changes, which can be prohibitive even under sparse changes in the hub nodes.  Estimating the difference of two precision matrices is a relatively easier task and has received attention recently \cite{zhao2014direct,yuan2017differential,jiang2018direct,tang2020fast}, providing finite-sample guarantees in the high-dimensional regime when such a difference is sparse. An existing study closest to the scope of our work is \cite{ghoshal2019direct}, which proposes re-estimating precision difference to progressively eliminate nodes and estimate the difference DAG. This approach critically hinges on the assumption that the noise variance is invariant, rendering limited applicability to intervention settings. In contrast, our algorithm builds on the changes in noise variances and estimates the intervention targets efficiently. We demonstrate the effect of this difference in Appendix \ref{sec:additional_intervention_recovery}.

\textbf{Learning interventional-MEC:} There is a growing number of studies on causal structure learning from both observational and interventional data. Score-based greedy interventional equivalence search (GIES) \cite{gies} and hybrid interventional greedy sparsest permutation (IGSP) \cite{igsp} algorithms are proposed for settings in which there are no latent confounders. Both of these algorithms assume that the intervention targets are known. However, knowing the intervention targets can be a strong assumption since even for controlled interventional experiments, off-target effects are common. For instance, noise variances of the off-target variables can change, resulting in the intervened nodes being unknown.

For causal structure learning without the knowledge of the intervention targets, an existing study includes the dynamic programming approach in \cite{eaton2007}. This approach has limited scalability due to its time complexity being exponential in the model size. Structural discovery from interventions proposed in \cite{ke2020learning} is a neural network-based method that can learn from interventional data without target knowledge. Nevertheless, it requires discrete data and can have at most one intervened node in a given setting, limiting its applicability. The differential causal discovery from interventional data (DCDI) algorithm proposed in \cite{DCDI} extends the approach of \cite{ke2020learning} without making as strong assumptions, e.g., learning a distribution over all potential interventional families via continuous optimization. While this method is shown to converge to the real intervention targets, its runtime becomes prohibiting even in models with as few as 100 nodes. The Unknown-target IGSP (UT-IGSP) algorithm proposed in \cite{utigsp} learns the intervention targets simultaneously while learning the causal structure. Even though interventions refine the search space, a greedy search of the sparsest permutation is still too slow in the high-dimensional regime, especially when using non-Gaussian CI tests. A graphical characterization of soft interventions with unknown targets is proposed in \cite{jaber2020causal} for causally insufficient systems. This algorithm, however, relies on CI tests and it is not scalable. We focus on causally sufficient systems in this paper.

\section{Problem definition}\label{sec:problem}

Let $\mcG \triangleq ([p],E)$ be a DAG with the node set $[p]\triangleq\{1,\dots,p\}$ and the edge set $E \subseteq [p] \times [p]$. We denote the directed edge from $i \in [p]$ to $j \in [p]$ by $i\rightarrow j$. We associate a random variable $X_i$ with $i \in [p]$, and accordingly, define the random vector $X \triangleq (X_1,\dots,X_p)^\top $. We consider a linear SEM, according to which
\begin{align}
    X = B^\top X+\epsilon \ ,
\end{align} 
where $B \in \mathbb{R}^{p \times p}$ is the autoregressive matrix in which $B_{ij}\neq 0$ if and only if $i \rightarrow j$ in $\mcG$. The random vector $\epsilon \in \mathbb{R}^{p\times 1}$ has zero-mean and covariance matrix $\Omega \triangleq {\sf diag}(\sigma_1^2,\dots,\sigma_p^2)$. We denote the covariance matrix of $X$ by $\Sigma$ and its inverse (the precision matrix) by $\Theta$, which satisfies $\Theta = (I-B)\Omega^{-1}(I-B)^\top$. For entries of $\Theta$, we have 
\begin{align}
    \Theta_{ij} &= -\frac{B_{ij}}{\sigma_j^{2}} - \frac{B_{ji}}{\sigma_i^{2}} + \sum_{k \in \Ch(i) \cap \Ch(j)} \frac{B_{ik} B_{jk}}{\sigma_k^2} \ , \quad \forall i \neq j \ , \label{eq:precision_ij_expression} \\
  \text{and} \quad  \Theta_{ii} &= \sigma_i^{-2} + \sum_{j \in \Ch(i)} \sigma_j^{-2} B_{ij}^2 \ , \quad \forall i \in [p] \ ,  \label{eq:precision_ii_expression}
\end{align}
where $\Ch(i),\Pa(i),\De(i)$, and $\An(i)$ denote children, parents, descendants, and ancestors set of node $i$ in DAG $\mcG$, respectively.

From the observational data, a DAG can be learned only up to its MEC \cite{verma1992algorithm}. Interventions are used to increase the identifiability of a DAG by removing all causes of the intervention target (\emph{perfect intervention}) or modifying those relationships without removing them completely (\emph{imperfect intervention}). We consider the following \emph{soft intervention} setting, which does not remove causal effects (from direct parents) on intervention target nodes, and hence, is more practical.

\noindent\textbf{Soft intervention model.}
In this model, interventions correspond to disturbing the target nodes $i \in \mcI$ by changing the variances of their noise variables, while the cause weights, i.e., the weights $B_{\Pa(i),i} \triangleq \{ B_{j,i}: j \in \Pa(i)\}$, can vary freely. This intervention procedure on the initial DAG results in a second DAG with new parameters.

Let $\mcGo$ represent a linear SEM prior to intervention with parameters $\Bo,\eo$, and $\mcGt$ be the linear SEM after the intervention, with parameters $\Bt,\et$. The intervention target set that relates these two DAGs is
\begin{align}
    \mcI \triangleq \{i: \sigmao_i \neq \sigmat_i \} \ .
\end{align}
Accordingly, denote the covariance and precision matrices of these two models by $\Sigmao,\Sigmat,\Thetao$, and $\Thetat$. Accordingly, denote the differences between the two models by $\Delta_B \triangleq \Bo-\Bt$ and $\Delta_\Theta \triangleq \Thetao-\Thetat$. For a subset of nodes $S \subseteq [p]$, denote the precision matrix of the random vector $X_S \triangleq \{X_i : i\in S\}$ by $\Theta_S$. We also denote the set of changed nodes by $S_\Delta \triangleq \{k: (\Delta_\Theta)_{k,k} \neq 0 \}$ and denote its size by $p_\Delta \triangleq |S_\Delta|$. According to \eqref{eq:precision_ii_expression}, this set consists of all the intervened nodes and their parents.

In this paper, we estimate the intervention targets $\mcI$ given the data from two SEMs under the soft intervention model. Furthermore, we estimate the non-intervened parents of the targets, which update the given observational MEC into the $\mcI$-MEC. Formally, we define~{$\psi: \Sigmao \times \Sigmat \rightarrow \left(\hat\mcI, \Pa(\hat\mcI)\right)$} as the estimator that maps the covariance matrices of the observational and interventional data to an intervention target set estimate and their parents. We aim to maximize the probabilities of $\psi$ recovers $\mcI$ and their parents. To this end, we define 
\begin{align}
    \P  \triangleq \mP(\mcI=\hat\mcI)   \qquad \text{and} \qquad \Q \triangleq \mP(\Pa(\mcI)=\Pa(\hat\mcI)) \ .
\end{align}
We will show that the algorithm proposed in Section \ref{sec:algorithm} exactly recovers $\mcI$ and the non-intervened parents of the members of $\mcI$. Hence, given the observational MEC, we obtain the $\mcI$-MEC.

\section{Algorithm and main results}\label{sec:algorithm}
In this section, we provide our proposed algorithm and present the attendant performance guarantees. Our algorithm involves repeatedly estimating the difference of the precision matrices to find the intervention target set $\mcI$, or equivalently, its complement $\mcI^\C$. This algorithm consists of three key steps. In \textbf{Step~1}, instead of directly estimating $\mcI$, we aim to discard the nodes that are strongly deemed not to belong to $\mcI$. For this purpose, we start by identifying the non-intervened nodes that do not have intervened children. These nodes are not of interest and they are discarded from further consideration. {We achieve this by estimating difference of precision matrices corresponding to the complete model with variables $[p]$, and we denote the changed nodes in the diagonal of this precision difference matrix with $S_\Delta$.} We continue with only the nodes contained in $S_\Delta$ for further scrutiny. A naive approach to identifying the rest of the non-intervened nodes is computing $\Delta_\Theta$ exhaustively for all the $2^{p_\Delta}$ subsets of $S_\Delta$, which for large $p_\Delta$ is computationally prohibitive. Alternatively, we partition  $S_\Delta$ into two sets: the set of  \emph{non-intervened source} nodes and its complement. We feed these two partitions for further processing to Step~2. Since the distribution of a node relies only on its ancestors, reaching a topological ordering is critical to reduce the complexity. In \textbf{Step 2}, for each source node, we find the nodes that share a common ancestor with it. Subsequently, we decompose $S_\Delta$ into equivalence classes according to these ancestral relationships with non-intervened source nodes. Decomposing $S_\Delta$ allows us to order the equivalence classes according to a topological ordering. In \textbf{Step~3}, we process these classes individually, and show that we only need to compute $\Delta_\Theta$ for all the subsets of each class considered in Step~3. This results in a significant reduction in the computational complexity compared to the exhaustive search approach. Finally, we identify the non-intervened parents of $\mcI$ from the earlier results. Since estimating the precision difference appears in our algorithm repeatedly, we describe it next and then continue with the details of the three steps.

\noindent\textbf{Precision difference estimation (PDE).}
When the difference between two SEMs is sparse, estimating the difference of their precision matrices can be formulated as a Lasso-type problem and solved efficiently. Since it will be used repeatedly, it is important for this function to recover the support of $\Delta_\Theta$ and have a feasible computational complexity. The study in \cite{jiang2018direct} solves the following convex problem through the alternating direction method of multipliers (ADMM) to estimate $\Delta_\Theta = \Thetao - \Thetat$: 
\begin{align}\label{eq:loss_daqda}
    \hat\Delta_\Theta = \argmin_{\Delta_\Theta} \left\{ \frac{1}{2}\sf{Tr}(\Delta_\Theta^\top\hat\Sigma^{(1)}\Delta_\Theta\hat\Sigma^{(2)}) - \sf{Tr}(\Delta_\Theta(\hat\Sigma^{(1)}-\hat\Sigma^{(2)})) + \lambda \|\Delta_\Theta\|_1 \right\} \ ,
\end{align}
where $\lambda$ is a tuning parameter. The computational complexity of this algorithm grows according to $O(p^3)$, which overcomes the limitation of the algorithm of \cite{zhao2014direct} used in \cite{ghoshal2019direct}, which has complexity $O(p^4)$. Therefore, we use this ADMM-based method as our PDE procedure.

\paragraph{Step 1: Finding the non-intervened source nodes.}By leveraging the PDE procedure discussed, we first estimate $\Delta_\Theta$ over all $[p]$ nodes and obtain $S_\Delta$. Representation of $\Theta_{i,i}$ in \eqref{eq:precision_ii_expression} implies that a diagonal entry of $\Delta_\Theta$ is non-zero if and only if either its corresponding node is in $\mcI$ or it has a child in $\mcI$. Hence, the set $[p]\setminus S_\Delta$ contains only non-intervened nodes and can be discarded from consideration. Furthermore, some non-intervened nodes in $S_\Delta$, namely those that do not have intervened ancestors, do not observe a change in their statistics. Therefore, it is possible to identify them from covariance matrices $\Sigmao$ and $\Sigmat$. Subsequently, these nodes can serve as the starting points for distinguishing the rest of the $\mcI^\C$ in~$S_\Delta$. Let us define them as the \emph{non-intervened source nodes}, denoted by 
\begin{align}\label{eq:J0}
    J_0 \triangleq \{j: j \in S_\Delta, \; j \notin \mcI, \; \An_\mcI(j) =\emptyset \} \ . 
\end{align}
The outputs of this step, $S_\Delta$ and $J_0$, are fed into the next steps of the algorithm.

\noindent\textbf{Step 2: Forming equivalence classes from $J_0$.}
We will show that for any non-intervened node $j \in S_\Delta$, there exists a minimal subset of $S_\Delta$, which makes the corresponding diagonal entry of the precision matrix invariant, and it does not contain any descendant of $j$. Therefore, the non-intervened nodes that have the same ancestors are affected by the interventions similarly, and finding their ancestors is critical. We show that determining whether node $k \in S_\Delta \setminus J_0$ has a common ancestor with node $j \in J_0$ is possible by applying PDE on $\{j,k\}$. Accordingly, we define the \emph{source ancestral set} $J_0^k$ for each node $k \in S_\Delta \setminus J_0$ as 
\begin{align}
    J_0^k & \triangleq \{j: j\in J_0, \;  \An(j) \cap \An(k) \neq \emptyset \}, \;\;\; \forall  k \in S_\Delta \setminus J_0 \ .  \label{eq:J0k} 
\end{align}
Next, we decompose the set $S_\Delta \setminus J_0$ into \emph{equivalence classes} where all the nodes in a class have the same source ancestral set. We denote these equivalence classes by $\mcA_1,\dots,\mcA_L$, and the source ancestral set corresponding to a class $\mcA_\ell$ by~$J_0^{\mcA_\ell}$ for $\ell \in [L]$. These classes are ordered according to a topological order such that for $ 1\leq \ell < \ell' \leq L$, $\smash{J_0^{\mcA_{\ell'}} \not \subset J_0^{\mcA_\ell}}$. In other words, the class corresponding to the superset of any $J_0^{\mcA_\ell}$ should appear later than $\mcA_\ell$ in the sequence $\mcA_1,\dots,\mcA_L$. Source ancestral sets and equivalence classes are fed into the next step.

\noindent\textbf{Step 3: Processing equivalence classes.}
Given Step 1 and Step 2, we can describe our exact search space of subsets for $\Delta_\Theta$ estimates to declare whether a node is intervened. We process equivalence classes $\mcA_1,\dots,\mcA_L$ individually, i.e., at stage $\ell$, we consider the nodes in $\mcA_\ell$. Let us define $\smash{\mcM_\ell \triangleq J_0 \cup \bigcup_{1\leq b < \ell}\mcA_b}$. It suffices to estimate $\smash{\Delta_{\Theta_{\mcM_\ell \cup A}}}$ \textbf{only for each $A \subseteq \mcA_\ell$} to determine the intervention status of any node in $\mcA_\ell$ class. This key observation reduces the number of PDE steps needed. Specifically, for any non-intervened $j \in \mcA_\ell$, there exists a subset $A \subseteq \mcA_\ell$ such that the corresponding diagonal entry of  $\smash{\Delta_{\Theta_{\mcM_\ell \cup A}}}$ will be zero, and there does not exist any such set for the intervened nodes in $\mcA_\ell$. Formally, the \emph{process equivalence class} returns
\begin{align}
    \mcI_\ell = \{i: i \in \mcA_\ell \cap \mcI\}\ , \quad \mbox{and} \quad
    J_\ell &= \{j : j \in \mcA_\ell \cap \mcI^\C \}\ .
\end{align}
Finally, we identify the non-intervened parents of the intervened nodes without any new $\Delta_\Theta$ estimates.

\noindent\textbf{Computational complexity.}
Algorithm \ref{alg:main_algorithm} repeatedly performs PDE in each step. The number of required instances of PDE is $(p_\Delta +1)$ in Step 1, $O(p_\Delta^2)$ in Step 2, and $\smash{\sum_{\ell \in [L]} 2^{|\mcA_\ell|}}$ in Step 3. Hence, it grows exponentially with $\smash{\max_{\ell \in [L]} |\mcA_\ell|}$, which can be $p_\Delta$ in the worst case in extreme examples. Nevertheless, in almost all practical scenarios it is usually considerably smaller. To provide some insights, we provide the next example and relegate more discussions to Appendix~\ref{sec:additional_computational_complexity}.

\begin{minipage}{0.7\linewidth}
\noindent\textbf{Example 1.}
Consider a DAG with nodes $\{1,2,3,4,5\}$ and the edge set $\{1 \rightarrow 3, 3 \rightarrow 4, 2\rightarrow 4, 2\rightarrow 5, 4 \rightarrow 5 \}$ and let $\mcI = \{3,5\}$. Hence, we have $S_\Delta=\{1,2,3,4,5\}, J_0=\{1,2\}, J_0^3 = \{1\}, J_0^4 = \{1,2\}, J_0^5= \{1,2\}$, and accordingly, $\mcA_1 = \{3\}, \mcA_2 =\{4,5\}, J_0^{\mcA_1}=\{1\}$, and  $J_0^{\mcA_2}=\{1,2\}$. Note that the largest $\mcA$ class has 2 nodes whereas $S_\Delta$ has 5 nodes.
\end{minipage}\hfill
\begin{minipage}{0.25\linewidth}\includegraphics[width=0.7\linewidth]{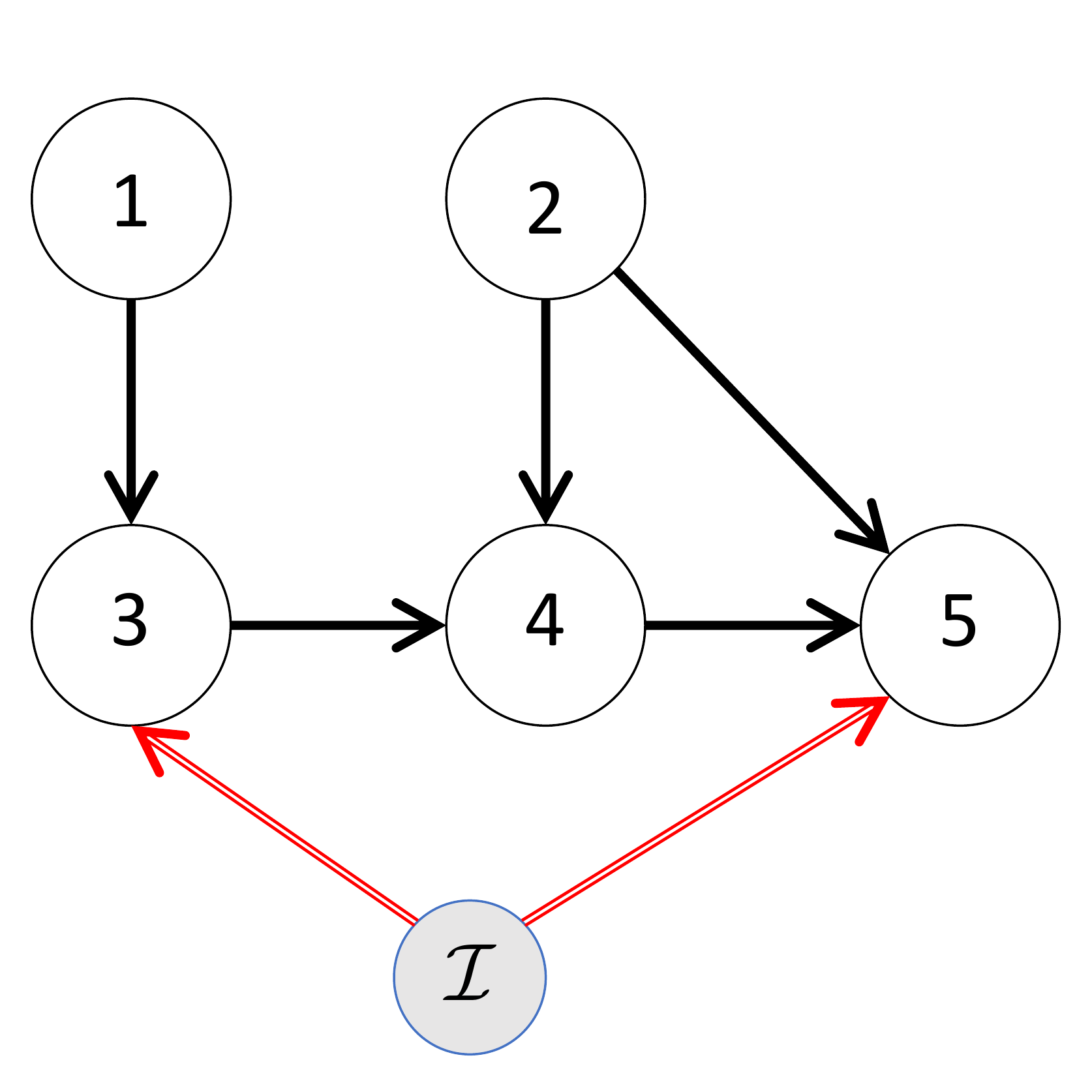}
\end{minipage}

\begin{algorithm}[h]
\caption{Causal Intervention Target Estimator (CITE)}
\label{alg:main_algorithm}
\begin{algorithmic}[1]
\State \textbf{Input:} ${\Sigma}^{(1)}$ and ${\Sigma}^{(2)}$
\State \textbf{Output.} $\mcI$ intervention set
\State Estimate $\Delta_\Theta \leftarrow \emph{precision difference} ({\Sigma}^{(1)},{\Sigma}^{(2)})$
\State Form $S_\Delta \triangleq \{k: k \in [p], \; [\Delta_\Theta]_{k,k}\neq 0\}$
\State Form $J_0$ using \eqref{eq:J0} and $J_0^k$ for each $k \in S_\Delta \setminus J_0$ using \eqref{eq:J0k}
\State Form equivalence classes $\mcA_1,\dots,\mcA_L$.
\For {$\ell \in [L]$}
    \State Take $\mcA_\ell$ set and the corresponding $J_0^{\mcA_\ell}$ set
    \State $\mcB_\ell \leftarrow \{b: J_0^{\mcA_b} \subset J_0^{\mcA_\ell}, \;\; 1\leq b < \ell\}$
    \State $\mcM_\ell = J_0^{\mcA_\ell} \cup \bigcup_{b \in \mcB_\ell}\mcA_b$
    \State $J_\ell, \mcI_\ell \leftarrow \emph{process equivalence class} (\mcM_\ell,\mcA_\ell,{\Sigma}^{(1)} ,{\Sigma}^{(2)})$ 
\EndFor
\State $\mcI = \bigcup_{\ell \in [L]} \mcI_\ell$
\State $\Pa(\mcI) \leftarrow \emph{parent finder} (\mcI,\mcJ,\mcM_1,\mcA_1,\dots,\mcM_L,\mcA_L,\Sigma^{(1)},\Sigma^{(2)}) $
\State \textbf{Return} $\mcI$ and $\Pa(\mcI)$ 
\end{algorithmic}
\end{algorithm}

\begin{algorithm}[h]
\caption{Functions for the main algorithm}
\label{alg:algo2}
\begin{algorithmic}[1]
\Algphase{Precision Difference Estimation (PDE) ($\Sigmao,\Sigmat$)}
\State Using ADMM based algorithm of \cite{jiang2018direct} method estimate $\Delta_\Theta = ({\Sigmao})^{-1}-({\Sigmat})^{-1}$
\State Symmetrize $\Delta_\Theta$: set $\Delta_\Theta = (\Delta_\Theta + \Delta_\Theta^\top)/2$
\State Threshold $\Delta_\Theta$: set $[\Delta_\Theta]_{i,j} = 0$ if $|[\Delta_\Theta]_{i,j}|<\varepsilon$.
\State \textbf{Return} $\Delta_\Theta$
\end{algorithmic}

\label{alg:algo3}
\begin{algorithmic}[1]
\Algphase{Process Equivalence Class ($\mcM, \mcA, \Sigmao, \; \Sigmat$)}
\State For each subset $A \subseteq \mcA$, estimate $ \Delta_{{\Theta}_{\mcM \cup A}} \leftarrow \emph{precision difference}((\Sigmao)_{A,A},(\Sigmat)_{A,A})$
\For{$k \in \mcA$}
\If {$\exists A \subseteq \mcA$, where $k \in A$, and $[\Delta_{{\Theta}_{\mcM \cup A}}]_{k,k}=0$}
\State $J \leftarrow J \cup k$
\Else
\State $\mcI \leftarrow \mcI \cup k$
\EndIf
\EndFor
\State \textbf{Return} $J,\mcI$ 
\end{algorithmic}

\label{alg:algo4}
\begin{algorithmic}[1]
\Algphase{Parent Finder ($\mcI,\mcJ,\mcM_1,\mcA_1,\dots,\mcM_L,\mcA_L,\Sigmao,\Sigmat$)}
\For{$i \in \mcI$}
\State $c_i \leftarrow c_i \triangleq \ell : i \in \mcA_\ell $
\For{$j \in \mcM_{c_i}$}
\If {$\nexists A \subseteq \mcA_{c_i}$ such that $[\Delta_{\Theta_{\mcM_{c_i} \cup A}}]_{j,i}=0$}
\State Add $j$ to $\Pa(i)$
\EndIf
\EndFor
\EndFor
\State \textbf{Return} $\Pa(i)$ for $i \in \mcI$ 
\end{algorithmic}
\end{algorithm}

\noindent\textbf{Restricted SEM.}
For a linear SEM $\mathcal{G}$ with $(B,\epsilon)$, we denote the restricted SEM (see Lemma \ref{eq:lemma1} for details) that characterizes the relationship among the random variables $X_S$ for a set $S$ by $(B_S,\epsilon_S)$. As defined earlier, the corresponding precision matrix is denoted by $\Theta_S$. The entries of $B_S$ and noise variances $\sigma_S$ are characterized by the original values of $B$, $\sigma$, and $\Theta$.
\begin{remark}\label{remark:logic}
We remark that the invariance of the distributions for the noise term of a node and the value of the node are equivalent only for the non-intervened nodes that do not have an intervened ancestor. Therefore, only such nodes can be detected from the full linear SEM. The noise term of a non-intervened node maintains its invariance in a restricted SEM in which we keep its intervened ancestors and their parents. On the other hand, the noise term of an intervened node is always variant for any choice of restricted SEM.
\end{remark}

In the following subsections, we will provide different analytical guarantees of Algorithm \ref{alg:main_algorithm}. Specifically, we will comment on the consistency of $\mcI$ recovery, the refinement of MEC to $\mcI$-MEC, and the sample complexity.

\subsection{Consistency of $\mcI$ recovery}
We provide the consistency of Algorithm \ref{alg:main_algorithm} for estimating $\mcI$ in this subsection. First, we need the following assumption to ensure that interventions are successful. 

\begin{assumption}[$\mcI$-faithfulness] \label{eq:assumption}
For any choice of $i,j \in S \subseteq [p]$, we have the following properties:
\begin{enumerate}
    \item If ${\sigmao_i} \neq \sigmat_i$, then $\sigma_{S,i}^{(1)} \neq \sigma_{S,i}^{(2)}$.
    \item If $\sigma_{S,i}^{(1)} \neq \sigma_{S,i}^{(2)}$, then $[\Theta_S^{(1)}]_{i,i}\neq [\Theta_S^{(2)}]_{i,i}$. 
    \item If $[B_S]_{j,i}\neq 0$ in either model, then   $[\Theta_S^{(1)}]_{i,j}\neq [\Theta_S^{(2)}]_{i,j}$.
\end{enumerate}
\end{assumption}
Next, we characterize the parameters of a restricted SEM, and then formalize the observations stated in Remark \ref{remark:logic} in the subsequent proposition.
\begin{lemma}[\cite{ghoshal2019direct}]\label{eq:lemma1}
Corresponding to a subset $S \subseteq [p]$, denote the removed set of nodes by $U \triangleq[p]\setminus S$ and define $U_j \triangleq U \cap \An(j)$, for $j \in S$. We have
\begin{align}
    \sigma_{S,j}^2 &= \sigma_j^4 \left(\sigma_j^2 - B_{U_j,j}^\top [\Theta_{\An(j)}]_{U_j,U_j}^{-1}B_{U_j,j} \right)^{-1} , \\
    [B_S]_{k,j} &= \frac{\sigma_{S,j}^2}{\sigma_j^2}\left(B_{k,j}-B_{U_j,j}^\top [\Theta_{\An(j)}]_{U_j,U_j}^{-1}[\Theta_{\An(j)}]_{U_j,k}\right) .
\end{align}
\end{lemma}

\begin{proposition}\label{prop:intervened_ancestors}
Denote the ancestors of $j\notin \mcI$ in $\mcI$ by $\An_\mcI(j)$. If a set $S$ contains $\An_\mcI(j)$ and their parents $\Pa(\An_\mcI(j))$, then $\sigma_{S,j}^{(1)}=\sigma_{S,j}^{(2)}$. Furthermore, for $i \in \mcI$ and any set $S$ we have $[\Delta_{\Theta_S}]_{i,i}\neq 0$. Additionally, if $[B_S]_{j,i} \neq 0$ in either model, then we have $[\Delta_{\Theta_S}]_{j,i}\neq 0$.
\end{proposition}

\begin{remark}
We repeatedly use the restricted SEM characterization in Lemma \ref{eq:lemma1} with various strategic choices of subsets $S$ in Algorithm \ref{alg:main_algorithm} to eliminate the non-intervened nodes from $S_{\Delta}$ using the criterion of Proposition \ref{prop:intervened_ancestors}. In Step 1, to identify $J_0$ in \eqref{eq:J0}, we set $S=\{j\}$ for each $j \in S_\Delta$. In Step 2, to identify $J_0^k$ in \eqref{eq:J0k} for each $k \in S_\Delta \setminus J_0$, we set $S=\{j,k\}$ for each $j \in J_0$. In Step 3, to process the nodes in $\mcA_\ell$, we use subsets of the form $\mcM_\ell \cup A$ for subsets $A$ in $\mcA_\ell$.
\end{remark}

\begin{theorem}[Consistency] \label{theorem:consistency}
Given Assumption \ref{eq:assumption} and the population covariance matrices, Algorithm~\ref{alg:main_algorithm} is consistent in estimating intervention target set $\mcI$ under soft interventions with $\P = 1$.
\end{theorem}

\subsection{\texorpdfstring{$\mcI$-Markov equivalence}{I-Markov equivalence}} \label{sec:I-Markov}
Interventions in a DAG change the conditional distributions of the intervened variables, and hence, they reveal orientations of some edges that were previously undirected in observational CPDAG, resulting in the interventional CPDAG ($\mcI$-CPDAG). The DAGs that have the same $\mcI$-CPDAG under soft intervention $\mcI$ form the $\mcI$-Markov equivalence class ($\mcI$-MEC). This is shown and discussed next. 

For a DAG $\mathcal{G}$ and an intervention set $\mcI$, an additional $\mcI$-vertex $\zeta$ and corresponding $\mcI$-edges $\{\zeta \rightarrow i\}_{i \in \mcI}$ are added to form the interventional DAG ($\mcI$-DAG). Note that vertex $\zeta$ creates a new v-structure $\zeta-i-j$ for any non-intervened $j \in \Pa(i)$. However, if $j$ is also in $\mcI$, then $\mcI$-DAG also contains the $\zeta \rightarrow j$ edge, and there is no new v-structure that can orient the edge $i-j$. 

We call the edges in $\mcG$ that are not directed in the original CPDAG but are directed in $\mcI$-CPDAG as $\mcI$-directed edges. In the \emph{parent finder} step of Algorithm \ref{alg:main_algorithm}, we find the edge set ~${\{j \rightarrow i\}_{j \notin \mcI, i \in \mcI}}$, and subsequently, obtain the $\mcI$-MEC. Therefore, we can use Algorithm \ref{alg:main_algorithm} in conjunction with an observational algorithm to perform causal structure learning, and establish the following theorem.
\begin{theorem}[$\mcI$-MEC] \label{theorem:parents}
Algorithm \ref{alg:main_algorithm} consistently recovers non-intervened parents of an $i \in \mcI$ with $\Q=1$ in population setting. This result modifies the original MEC, which can be obtained via any observational structure learning  algorithm, into the $\mcI$-MEC.
\end{theorem}

\subsection{Sample complexity} \label{sec:sample_complexity}
In this subsection, we provide the finite-sample counterparts of Theorem \ref{theorem:consistency} and Theorem \ref{theorem:parents}. Our choice of the PDE algorithm, the ADMM-based method of \cite{jiang2018direct}, enjoys finite-sample results when noise $\epsilon$ has a Gaussian distribution. The following theorem establishes the sample complexity of Algorithm \ref{alg:main_algorithm} for estimating $\mcI$ and the non-intervened parents of the nodes in $\mcI$.

\begin{theorem}[Sample complexity] \label{theorem:sample_complexity}
Let $d$ denote the maximum degree of an intervened node and set $\Gamma \triangleq \Sigmat \otimes \Sigmao$ and  $\alpha \triangleq 1 - \max_{e \notin \supp} |\Gamma_{e,\supp}\Gamma_{\supp,\supp}^{-1}|_1$. Accordingly, define $M \triangleq \max\{ \| \Sigma^{(1)}\|_\infty, \|\Sigma^{(2)} \|_\infty\ \}$, $ M_\Sigma \triangleq  \max\{ \| \Sigma^{(1)} \|_{1,\infty}, \| \Sigma^{(2)} \|_{1,\infty} \}$, $M_{\Gamma,\Gamma^T} \triangleq \max\{\| \Gamma_{S,S}  \|_{1,\infty},\| \Gamma^T_{S,S}  \|_{1,\infty}\}$, where $S$ is the support of $(\Sigmat)^{-1} - (\Sigmao)^{-1}$.
When $\alpha >0$ and $M_\Sigma M_{\Gamma,\Gamma^T} < +\infty$, with $ n = O\left(\frac{d^4}{\varepsilon^2}\frac{\log p}{\delta}\right)$ samples, Algorithm \ref{alg:main_algorithm} 
\begin{enumerate}
    \item identifies $\mcI$ with a probability at least $\geq 1-\delta$; 
    \item identifies the non-intervened parents $\{j \rightarrow i\}_{j \notin \mcI, i \in \mcI}$ with a probability at least $ \geq 1-\delta$.
\end{enumerate}
\end{theorem}
Note that we have assumed that the product $M_\Sigma M_{\Gamma,\Gamma^T}$ is bounded. This is necessary to avoid a linear scaling of the sample complexity in $p$. More discussion on the necessity and implications of this assumption is provided in the proof of Theorem \ref{theorem:sample_complexity} in Appendix \ref{appendix:proofs}. 

\section{Empirical results}\label{sec:simulations}
\subsection{Synthetic data - intervention recovery} \label{sec:intervention_recovery}
We start by testing our algorithm for estimating  intervention targets, i.e., the set $\mcI$ . We generate 100 realizations of Erd\H{o}s-R\'enyi \cite{ErdosRenyi} DAGs with expected neighborhood size $c=1.5$, and $|\mcI|=5$. We sample the entries of $B$, i.e., the edge weights, independently at random according to the uniform distribution on $[-1,-0.25] \cup [0.25,1]$. The additive Gaussian noise terms have distribution $\mcN(0,I_p)$. We select the intervention set $\mcI$ by randomly selecting 5 nodes from $[p]$. We consider three different models to intervene on the nodes in $\mcI$: (i) \emph{shift intervention} model in which mean of the noise $\epsilon_i$ is shifted from 0 to 1, (ii) \emph{variance increase} model in which the variance of the noise $\epsilon_i$ is increased from 1 to 2, and (iii) \emph{randomized intervention} model, in which $(B^{(2)})_{\Pa(i),i}=0$ and the noise variance varies from 1 to 1.5. All the simulations are run on a MacBook Pro with 2.7 GHz Dual-Core i5 core and 8 GB RAM.

\begin{figure}[t]
    \centering
    \begin{subfigure}{0.32\textwidth}
        \centering
        \includegraphics[width=1\linewidth]{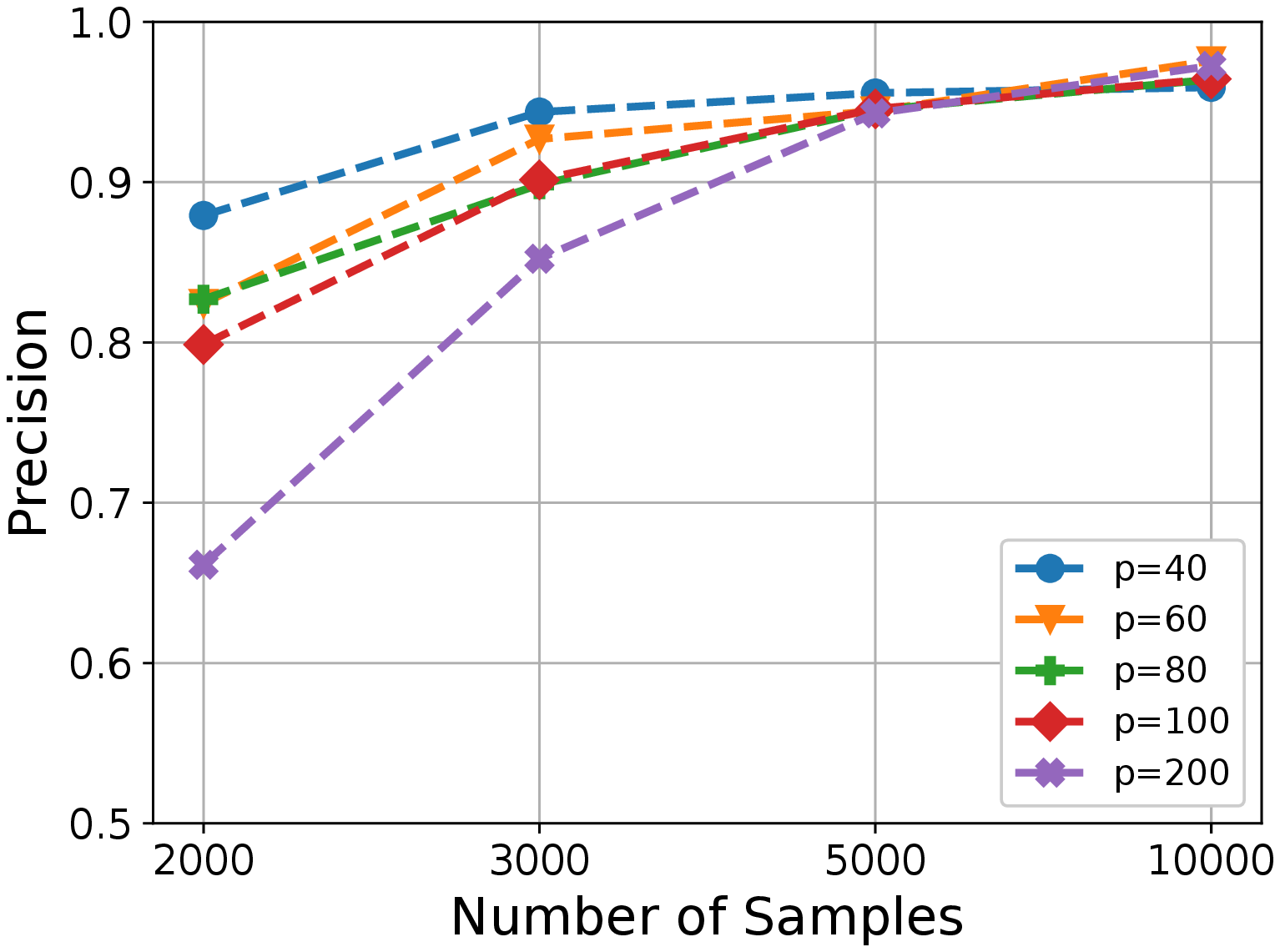}
        \caption{Shift intervention}
        \label{fig:ours_alone_shift}
    \end{subfigure}
    \begin{subfigure}{0.32\textwidth}
        \centering
        \includegraphics[width=1\linewidth]{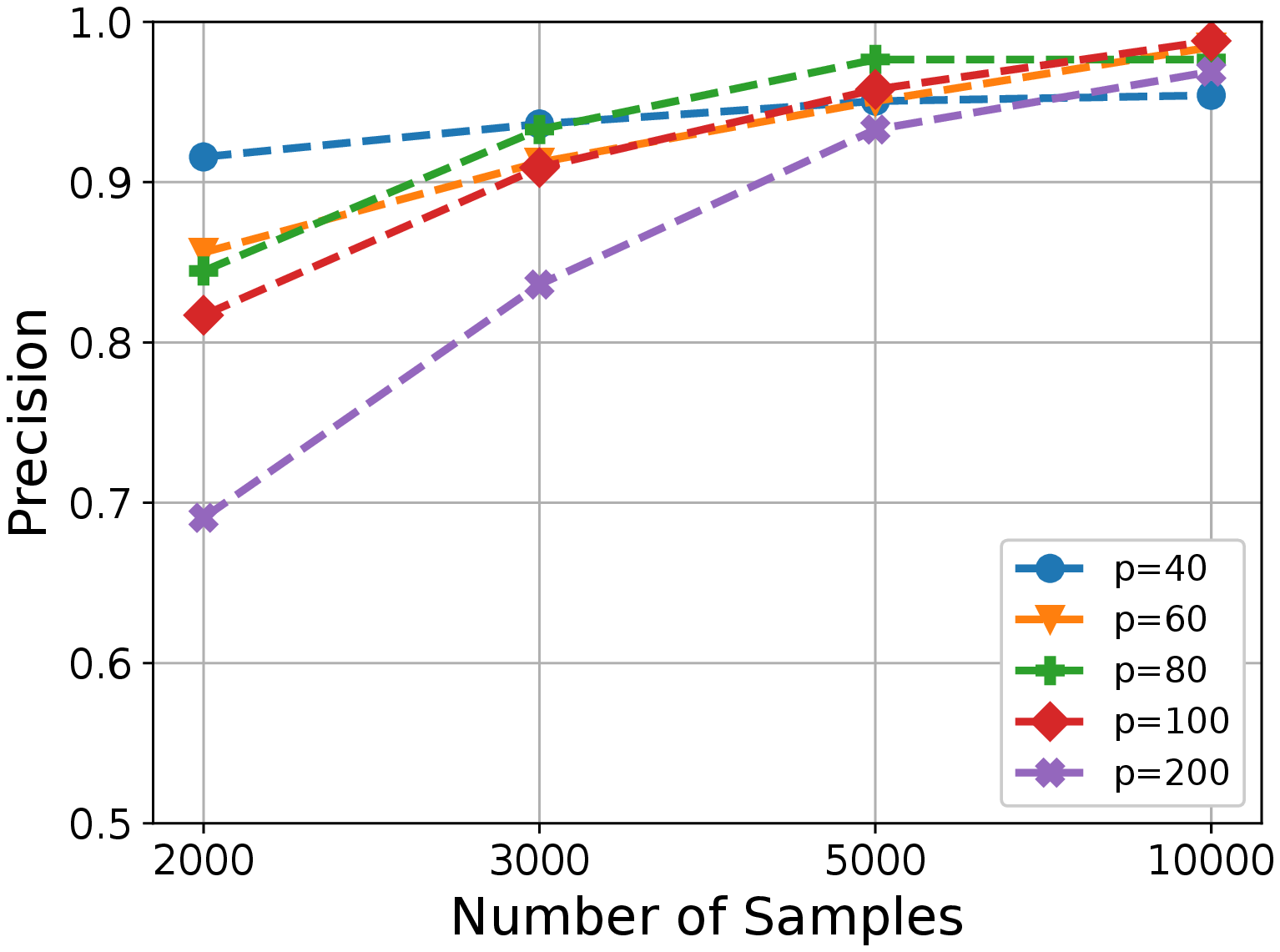}
        \caption{Increased variance}
        \label{fig:ours_alone_inc}
    \end{subfigure}    
    \begin{subfigure}{0.32\textwidth}
        \centering
        \includegraphics[width=1\linewidth]{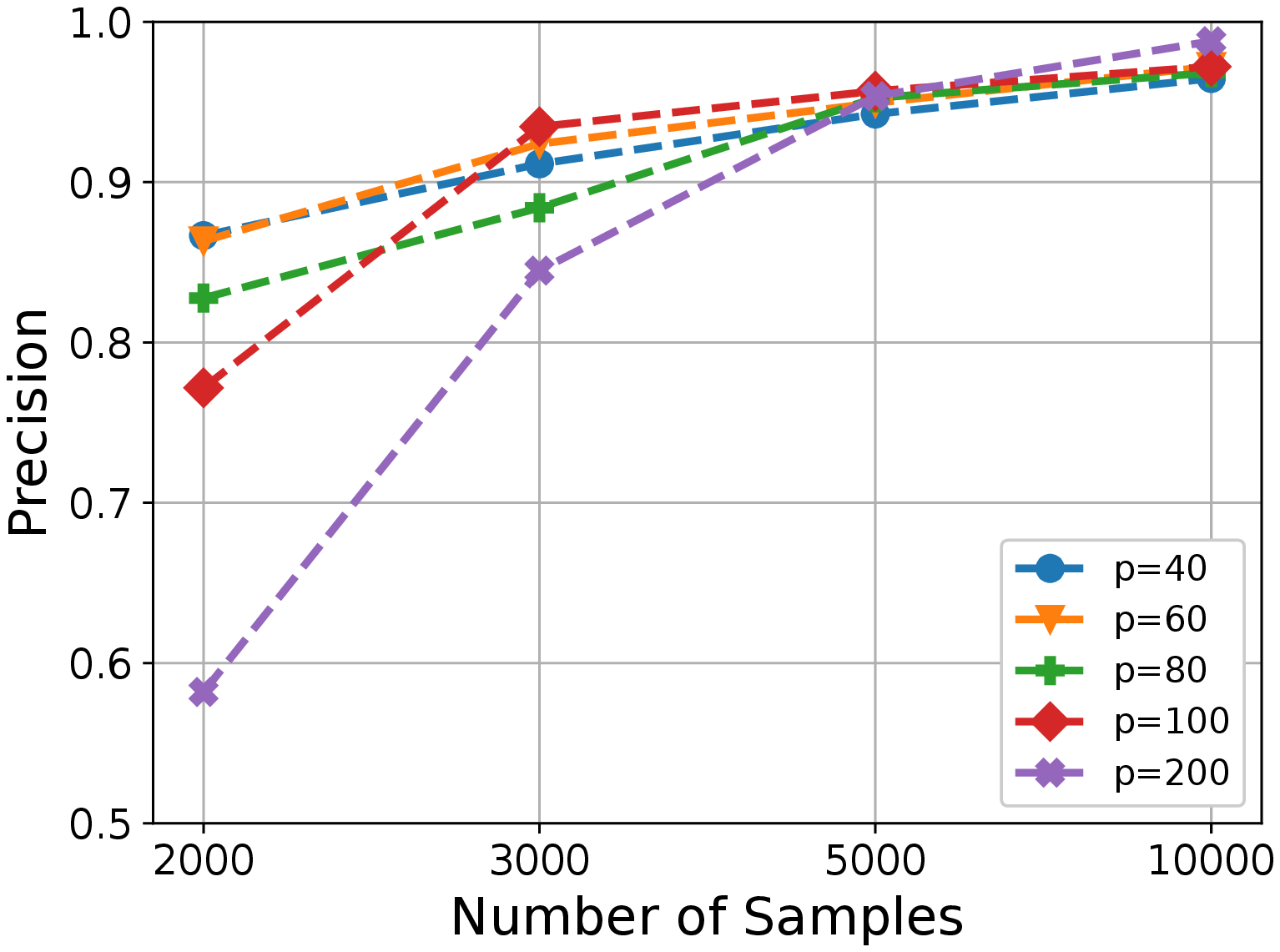}
        \caption{Randomized intervention}
        \label{fig:ours_alone_per}
    \end{subfigure}   
    \caption{Average precision of estimating intervention targets. Algorithm \ref{alg:main_algorithm} reaches high precision with increasing number of samples even in large models for all settings.}
    \label{fig:ours_alone}
\end{figure}

We first run our algorithm by varying the graph size $p$ and the number of samples. Figure \ref{fig:ours_alone} illustrates that our algorithm is able to recover the intervention targets with high precision under all three intervention models. Having high precision is especially important in high dimensions, since a large false positive rate severely affects any downstream task such as structure learning. Recall rates are close to 1 and they are omitted from the graph.  


Next, we compare our results with that of the UT-IGSP algorithm \cite{utigsp} for the shift intervention model. We note that UT-IGSP performs a greedy search to identify the sparsest permutation through CI tests, and it returns intervention targets as a by-product along with the learned causal structure. The computation time of UT-IGSP, hence, grows quickly with the size of the graph, reaching an average of 61.2 seconds for $p=100$. Therefore, the complexity of structure learning and intervention target discovery in the high-dimensional regime is prohibitive. In contrast, our algorithm has comparable performance to UT-IGSP when $p=100$, while requiring less than a second of runtime. Our algorithm's runtime scales gracefully when the dimension is in the hundreds. 

\begin{table}[h]
  \caption{$\mcI$ estimation in the shift intervention model - 50 repetitions with 5000 samples - density 1.5}
  \label{tab:shift}
  \centering
  \begin{tabular}{lllllll}
    \toprule
    \multicolumn{4}{c}{UT-IGSP (\cite{utigsp})} &
    \multicolumn{3}{c}{Algorithm \ref{alg:main_algorithm}}                   \\
    \cmidrule(r){1-7}
    p     & Precision & Recall & Time(s) & Precision & Recall & Time(s) \\
    \midrule
    40 & 0.99 (0.04)  & 0.99 (0.04) & 0.8 & 0.96 (0.09) & 0.94 (0.09) & 0.1    \\
    60 & 0.95 (0.07)  & 0.99 (0.05) & 5.2 & 0.97 (0.07) & 0.95 (0.10) & 0.2     \\
    80 & 0.96 (0.08)  & 0.99 (0.04) & 17.8 & 0.96 (0.08) & 0.96 (0.10) & 0.3     \\
    100 & 0.93 (0.11) & 1 (0) & 61.2 & 0.94 (0.09) & 0.98 (0.07) & 0.3 \\
    \bottomrule
  \end{tabular}
\end{table}

\subsection{Synthetic data - causal structure learning}
In Section \ref{sec:I-Markov}, we have shown that Algorithm \ref{alg:main_algorithm} recovers the intervention targets. It can be further used to refine the observational MEC to an $\mcI$-MEC. Accordingly, we take the correct CPDAG of $\mathcal{G}^{(1)}$, and then apply our algorithm's findings to obtain $\mcI$-CPDAG. We report the accuracy of additional edge orientations and in particular recovery of parents (if possible) of intervention targets in Appendix~\ref{sec:additional_causal_structure}.

\subsection{Application to real data}\label{sec:real_data}
We apply our algorithm to two real datasets with observational and interventional data to learn their causal structures. When there exist multiple interventional environments, we apply our algorithm to them individually to estimate the intervened nodes and their parents. Subsequently, we combine the results from all environments in order to form the final estimated structure. There is a large number of interventional settings in which finding the targets and their non-intervened parents by our algorithm yields a good estimate of the associated DAG, which we use for our evaluation.

We compare our results with those of the algorithms UT-IGSP, and UT-IGSP* in \cite{utigsp}\footnote{The code and preprocessed real data are taken from \hyperlink{https://github.com/csquires/utigsp}{https://github.com/csquires/utigsp} for fair comparison, and CausalDAG package which is distributed under 3-Clause BSD licence is used.}, where the former works with partially known intervention targets and the latter does not require any target input. We use both parametric (Gaussian) and non-parametric (Hilbert-Schmidt independence criterion) CI tests for UT-IGSP methods. We note that the non-parametric tests have significant runtimes. We note that our algorithm uses PDE at several stages, which calls for different $\lambda$ regularization parameters. Namely, let us denote the parameters used for Step 1 and Step 3 by $\lambda_1$, Step~2 by $\lambda_2$ and the \emph{parent finder} of Step 3 by $\lambda_3$. Similarly, UT-IGSP needs a cut-off value $\alpha$ for CI tests. We run the algorithms with different values of these parameters to obtain the receiver operating characteristic (ROC) curves.

\begin{figure}[t]
    \centering
    \begin{subfigure}{0.4\linewidth}
        \centering
        \includegraphics[width=1\linewidth]{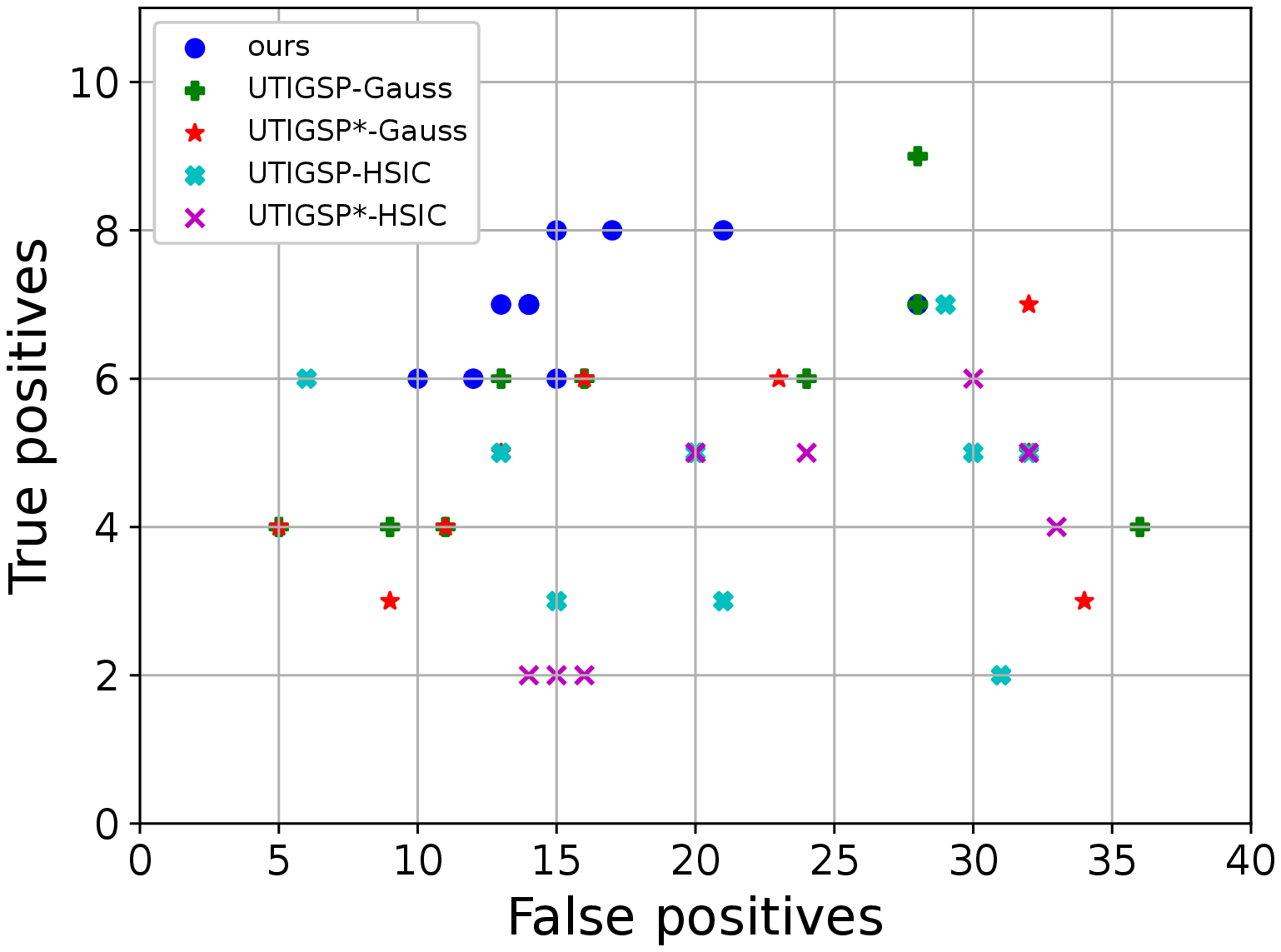}
        \caption{Sachs \cite{SACHSKaren2005Cpnd} dataset}
        \label{fig:sachs_directed}
    \end{subfigure}
    \begin{subfigure}{0.4\linewidth}
        \centering
        \includegraphics[width=1\linewidth]{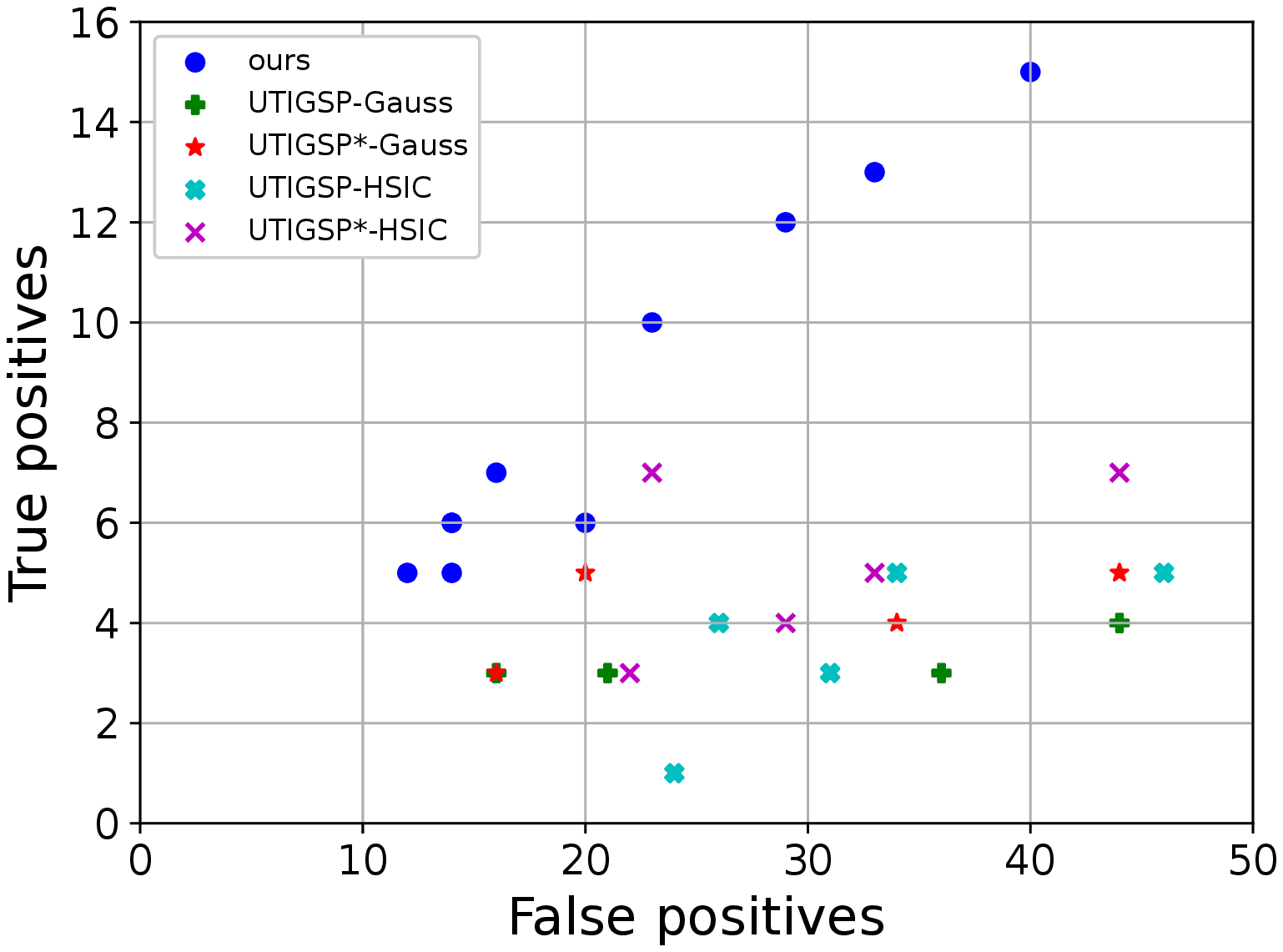}
        \caption{Dixit \cite{perturb-seq} dataset}
        \label{fig:dixit_directed}
    \end{subfigure}    
    \caption{ROC curves for directed edge recovery. Algorithm \ref{alg:main_algorithm}: (a) is more robust in smaller graphs; (b) handles larger graphs and multiple interventional settings successfully, and conforms to real datasets better than CI testers.}
    \label{fig:real_directed}
\end{figure}

\noindent\textbf{Protein signaling data.}
We first consider the dataset in \cite{SACHSKaren2005Cpnd} for discovering the protein signaling network of 11 nodes. It consists of measurements of proteins and phospolipids under different interventional environments. In each environment, signaling nodes are inhibited or activated. Hence, these sites form intervention targets. The conventionally accepted ground truth has been updated over the years, and we compare with the recent version in \cite{nesssachs2017}, which consists of 16 edges. We follow the process of \cite{utigsp} and work with 1755 observational and 4091 interventional samples aggregated from 5 different interventional environments. In Fig.~\ref{fig:sachs_directed}, we report the results of running Algorithm \ref{alg:main_algorithm} and UT-IGSP with various parameters.

\noindent\textbf{Perturb-seq gene expression data.}
We analyze the performance of our algorithm on the perturb-seq dataset by in \cite{perturb-seq}. The dataset consists of observational data and the interventional data from bone marrow-derived dendritic cells (BMDCs). A single gene has been targeted for deletion in each interventional environment. Similarly to \cite{perturb-seq}, we have focused on 24 genes that are known to have regulatory effects, and we have followed \cite{igsp} to select interventional data from 8 gene deletions along with observational samples. 
We take the Fig. 4D in \cite{perturb-seq} as the ground truth, which has 34 edges among 24 nodes. We use 23 interventional settings for the targeted 8 gene deletions. In Fig.~\ref{fig:dixit_directed}, we plot the results of running Algorithm \ref{alg:main_algorithm} and UT-IGSP with various parameters.

In both real datasets, our algorithm achieves higher accuracy in recovering directed edges. The comparison with UT-IGSP is more striking in Fig. \ref{fig:dixit_directed}, and shows that our ability to work with many interventional environments on a relatively larger graph. Furthermore, Fig. \ref{fig:sachs_directed} shows that Algorithm~\ref{alg:main_algorithm} handles smaller graphs more robustly. These results illustrate that even though our algorithm is designed for linear models, on real datasets, it performs better than the current state-of-the-art methods that rely on CI tests. 

\section{Conclusion}\label{sec:conclusion}
In this paper, we have considered the problem of estimating intervention targets in linear structural equation models (SEMs) under soft interventions. We have proposed an algorithm that consistently identifies intervened nodes that can scale to larger graphs and have sample complexity guarantees in Gaussian linear SEMs. The algorithm can be used also to infer interventional Markov equivalence class (MEC) from the observational MEC. We have demonstrated comparable or better performance compared to the existing methods in a number of settings.

The limitation of our method is that it only applies to linear SEMs. The dataset in an application should be evaluated carefully to confirm whether the assumptions are satisfied. This avoids any adverse effects arising from wrong interpretations of cause-effect relationships. Extending the similar ideas for scalable and efficient intervention target estimation to the non-linear DAGs is an open question that we aim to address in future work. Finally, the condition number of the optimization problem is assumed to be bounded in the sample complexity results. We note that our algorithm is independent of the specific precision difference estimation (PDE) algorithms and can be used in a modular way. In this regard, it can benefit from any potential relaxation on this limitation of PDE algorithms.

 

\bibliographystyle{IEEEtran}
\bibliography{references}

\newpage

\appendix

 {\centering \Large \textbf{"Scalable Intervention Target Estimation in Linear Models" \\ \hspace{1.6in} Supplementary Material} }

\section{Theoretical analysis}\label{appendix:proofs}

\paragraph{{\bf Proof of Lemma \ref{eq:lemma1}.}} This lemma is proved in \cite{ghoshal2019direct}. We provide an alternative and simplified proof for completeness, and with an approach that fits our subsequent analysis. The first observation is that the noise variance of a terminal node $j$ is the inverse of the corresponding diagonal entry of the precision matrix obtained by removing all descendants of $j$. Thus, if $j$ has no descendants in a set $S$, then $\sigma_{S,j}^{-2} = [\Theta_S]_{j,j}$. The second observation is that the noise variance of a node in a restricted SEM is affected by only its ancestors. Therefore, the noise variance of a node $j$ in a restricted SEM over $S$ is equal to that of over a set $S \cap \An(j)$, i.e., $\sigma_{S,j} = \sigma_{{S \cap \An(j)},j}$. Due to the second observation, we will consider only the restricted SEMs over the sets of the form $S = \An(j) \setminus U_j$, where $U_j$ denotes the ancestors of $j$ that lie out of this restricted SEM. Let us denote the precision matrix of the restricted SEM over $\An(j)$ by $\Phi \triangleq \Theta_{\An(j)}$. We obtain the variance of the noise term for a node $j \in S$ as follows:
\begin{align}
    \Phi_S &= \Phi_{S,S} - \Phi_{S,U_j}(\Phi_{U_j,U_j})^{-1}\Phi_{U_j,S} \ , \\
    \frac{1}{\sigma_{S,j}^2} &= \Phi_{j,j} = \frac{1}{\sigma_j^2} - \Phi_{j,U_j}(\Phi_{U_j,U_j})^{-1}\Phi_{U_j,j} \  \\
    &= \frac{1}{\sigma_j^2} - \frac{B_{U_j,j}^\top[\Theta_{\An(j)}]_{U_j,U_j}^{-1}B_{U_j,j}}{\sigma_j^4} \ ,
\end{align}
where the first line is due to Schur's complement; the second line is due to the observations mentioned above since $j$ is a terminal node in both sets $S$ and $\An(j)$; and the last line is due to \eqref{eq:precision_ij_expression} and Proposition~4 of \cite{ghoshal2018learning}.

We use the Markov property to characterize edge weights in a restricted SEM. Conditioned on all of its parents, node $j$ is independent of the remaining nodes. Hence, $[B_{\An(j)}]_{k,j} = B_{k,j}$. We consider the same set $S=\An(j) \setminus U_j$, $\Phi=\Theta_{\An(j)}$ and derive the edge weights as follows:
\begin{align}
    [\Phi_S]_{j,k} &= \Phi_{j,k} - \Phi_{j,U_j}(\Phi_{U_j,U_j})^{-1}\Phi_{U_j,k} \  \\ 
    & = -\frac{[B_{\An(j)}]_{k,j}}{\sigma_{\An(j),j}^2} + \frac{[B_{\An(j)}]_{U_j,j}^\top}{\sigma_{\An(j),j}^2}[\Theta_{\An(j)}]_{U_j,U_j}^{-1}[\Theta_{\An(j)}]_{U_j,k} \  \\
    &= -\frac{B_{k,j}}{\sigma_j^2} + \frac{B_{U_j,j}^\top}{\sigma_j^2}[\Theta_{\An(j)}]_{U_j,U_j}^{-1}[\Theta_{\An(j)}]_{U_j,k} \ , \\
    B^S_{k,j} & = \frac{\sigma_{S,j}^2}{\sigma_j^2}(B_{k,j} - B_{U_j,j}^\top [\Theta_{\An(j)}]_{U_j,U_j}^{-1}[\Theta_{\An(j)}]_{U_j,k}) \ ,
\end{align}
where the last line follows from $[\Phi_S]_{j,k} = -[B_S]_{k,j}/\sigma_{S,j}^2$.{ Note that this last equality is correct since $S$ contains only the ancestors of $j$. Similarly, we can write $[\Theta_S]_{j,j}=1/\sigma_{S,j}^2$ if $S$ contains only the ancestors of $j$. }
\endproof

\paragraph{{\bf Proof of Proposition \ref{prop:intervened_ancestors}.}} Let us consider the restricted SEM over set $S$ and let $U_j = \An(j) \setminus S$ denote the ancestors of $j$ that are not included in the restricted SEM. Note that the restricted SEM over $\An(j)$ has edge weights $B_{\An(j)} = [B]_{\An(j),\An(j)}$ and noise covariance $\Omega_{\An(j)} = [\Omega]_{\An(j),\An(j)}$. Therefore, for nodes $u,v \in U_j$, we can use Lemma~\ref{eq:lemma1} to obtain,
\begin{align}
 [\Theta_{\An(j)}]_{u,v} &= -\frac{[B_{\An(j)}]_{u,v}}{\sigma_{\An(j),v}^2}-\frac{[B_{\An(j)}]_{v,u}}{\sigma_{\An(j),u}^2}+\sum_{l \in \An(j)} \frac{[B_{\An(j)}]_{u,l}[B_{\An(j)}]_{v,l}}{\sigma_{\An(j),l}^2} \  \\
 &= -\frac{B_{u,v}}{\sigma_v^2}-\frac{B_{v,u}}{\sigma_u^2}+\sum_{l \in \An(j)} \frac{B_{u,l}B_{v,l}}{\sigma_l^2} \ , \label{eq:prop1_proof_l2} \\
 [\Theta_{\An(j)}]_{u,u} &= \frac{1}{\sigma_{\An(j),u}^2} + \sum_{l \in \An(j)}\frac{[B_{\An(j)}]_{u,l}^2}{\sigma_{\An(j),l}^2} \  \\
 &= \frac{1}{\sigma_u^2} + \sum_{l \in \An(j)} \frac{B_{u,l}^2}{\sigma_l^2} \ . \label{eq:prop1_proof_l4}
\end{align}
Now we will prove the first statement. If $S$ contains $\An_{\mcI}(j)$ and their parents $\Pa(\An_{\mcI}(j))$, we know that neither $u,v$ nor their children belong to $\mcI$. Therefore, $[\Theta_{\An(j)}]_{u,v}$ and $[\Theta_{\An(j)}]_{u,u}$ are invariant due to \eqref{eq:prop1_proof_l2} and \eqref{eq:prop1_proof_l4}, respectively. 
Subsequently, we have $[\Delta_{\Theta_{\An(j)}}]_{U_j,U_j}=0$. Furthermore, since $j \notin \mcI$, we have that $[\Delta_B]_{k,j}=0$ for $k\in [p]$. Using the Lemma~\ref{eq:lemma1} again, we obtain
\begin{align}
       \sigma_{S,j}^2 &= \sigma_j^2 \left(\sigma_j^4 - B_{U_j,j}^\top[\Theta_{\An(j)}]_{U_j,U_j}^{-1} B_{U_j,j} \right)^{-1},
\end{align}
where we note that $\sigma_j$, $B_{U_j,j}$, and $[\Theta_{\An(j)}]_{U_j,U_j}$ are all invariant and, subsequently, $\sigma_{S,j}^{(1)}=\sigma_{S,j}^{(2)}$ is invariant. This proves the first statement regarding the invariance of the noise term for a non-intervened node under certain restricted SEMs. 

For the last part, Assumption \ref{eq:assumption} ensures that $\sigma_{S,i}^{(1)}\neq \sigma_{S,i}^{(2)}$ for $i \in \mcI$, and we have $[\Delta_{\Theta_S}]_{i,i}\neq 0$. Similarly, Assumption \ref{eq:assumption} states that if $[B_S]_{j,i} \neq 0$ for either model, then $[\Delta_{\Theta_S}]_{j,i}\neq 0$.
 \endproof

 \paragraph{{\bf Proof of Theorem \ref{theorem:consistency}.}}
 We will follow the steps of the Algorithm \ref{alg:main_algorithm} to obtain the consistency results. We assume that the population-level statistics are known, i.e., we can compute $\Delta_{\Theta_S}$ for any $S \subseteq [p]$ correctly. Instead of estimating $\mcI$ directly, we, equivalently, aim to identify its complement $\mcI^\C$.

\emph{Forming $S_\Delta$.} In Step 1, we first estimate $\Delta_\Theta$ over $[p]$ to obtain the nodes that are affected by the interventions. Note that $\sigma_i^{(1)}\neq \sigma_i^{(2)}$ for intervened nodes $i \in \mcI$ and $[\Delta_B]_{k,j}=0$ for non-intervened nodes $j \notin \mcI$ and $k \in [p]$. According to  \eqref{eq:precision_ij_expression} and \eqref{eq:precision_ii_expression}, $[\Delta_B]_{k,k} \neq 0$ if and only if either $k \in \mcI$ or there exists $k\rightarrow i$ for which $i \in \mcI$. In other words, by forming the set $S_\Delta = \{k : k \in [p], [\Delta_\Theta]_{k,k} \neq 0\} = \mcI \cup \bigcup_{i \in \mcI} \Pa(i)$, we can discard the nodes in $[p]\setminus S_\Delta$. The discarded nodes consist of the non-intervened nodes that do not have children in $\mcI$. Next, we will show computationally, some of the non-intervened nodes in $S_\Delta$ can be identified  easier than the others. 

\emph{Forming non-intervened source nodes $J_0$.} Note that if a node $j$ has an intervened ancestor, the distribution of $X_j$ changes and, subsequently, $\Sigmao_{j,j} \neq \Sigmat_{j,j}$. If a node $i$ is intervened, the distribution of $X_i$ changes too, and it results in $\Sigmao_{i,i} \neq \Sigmat_{i,i}$. Therefore, we are able to find non-intervened source nodes directly from $\Sigmao$ and $\Sigmat$. Since we have already narrowed down our focus to set $S_\Delta$, we define \emph{non-intervened source nodes} as 
\begin{align}
 J_0  &\triangleq  \{j: j \in S_\Delta, \; j \notin \mcI, \; \An_{\mcI}(j) = \emptyset\} \  \\
 &= \{j: j \in S_\Delta, \;  \Sigmao_{j,j} = \Sigmat_{j,j} \} \ .
\end{align}
Sets $S_\Delta$ and $J_0$ are subsequently fed into the next steps of the algorithm.

\emph{Forming source ancestral sets $J_0^k$.}
In Proposition~\ref{prop:intervened_ancestors} we have shown that for any non-intervened node $j$, there exists sets $S$ that makes $\sigma_{S,j}$ invariant, and the condition is closely related to ancestors of $j$ that are affected by the intervention being included in $S$. On the other hand, such a restricted SEM does not exist for any intervened node. Therefore, we can identify all the non-intervened nodes in $S_\Delta \setminus J_0$ by finding a proper restricted SEM over a subset of $S_\Delta$. Hence, finding the ancestors of non-intervened nodes is critical. Now consider pair $\{j,k\}$ such that $j \in J_0$, $k \in S_\Delta \setminus J_0$. $\Sigma_{j,j}$ is invariant and $\Sigma_{k,k}$ is changing. If $j$ and $k$ have a common ancestor, which can be $j$ itself, then $\Sigma_{j,k}$ is nonzero and $[\Delta_{\Theta_{\{j,k\}}}]_{j,k}\neq 0$. Otherwise, $\Sigma_{j,k} = 0$ and  $[\Delta_{\Theta_{\{j,k\}}}]_{j,k} = 0$. Subsequently, we define the \emph{source ancestral set} $J_0^k$ for each node $k \in S_\Delta \setminus J_0$, that consists of the nodes in $J_0$ that have a common ancestor with $k$, i.e.,
\begin{align}
    J_0^k &\triangleq \{j: j \in J_0, \; [\Delta_{\Theta_{\{j,k\}}}]_{j,k}\neq 0\} \ , \;\;\; \forall  k \in S_\Delta \setminus J_0 \nonumber \  \\
    &= \{j: j\in J_0, \;  \An(j) \cap \An(k) \neq \emptyset \} \ .
\end{align}

Next, we will use these source ancestral sets to group the nodes that have similar ancestors together. 

\emph{Forming equivalence classes from $J_0$.}
We note that some of the nodes in $S_\Delta \setminus J_0$ will have identical {source ancestral sets}. Therefore, we can decompose the set $S_\Delta \setminus J_0$ into \emph{equivalence classes} such that all the nodes in a class have the same source {ancestral} sets. We denote these equivalence classes by $\mcA_1,\dots,\mcA_L$, and the source ancestral set corresponding to the class $\mcA_\ell$ by $J_0^{\mcA_\ell}$ for $\ell \in [L]$. Formally, 
\begin{align}
    S_\Delta \setminus J_0 &= \bigcup_{\ell \in [L]} \mcA_\ell \ , \\
    \mcA_{\ell_1} \cap \mcA_{\ell_2} &= \emptyset \ , \;\; \text{for} \;\;  \ell_1 \neq \ell_2 \ , \\
    J_0^{\mcA_\ell} \triangleq J_0^{k_1} &= J_0^{k_2},  \quad \forall k_1, k_2 \in \mcA_\ell \ , \;\; \text{for} \;\; \ell \in [L] \ . 
\end{align}

We note that we order these classes according to a topological order such that for $1\leq \ell < \ell' \leq L$, $J_0^{\mcA_{\ell'}} \not \subset J_0^{\mcA_\ell}$. In other words, the class corresponding to the superset of any $J_0^{\mcA_\ell}$ should appear after $\mcA_\ell$ in the sequence $\mcA_1,\dots,\mcA_L$. This ordering is important since we do not need descendants of a non-intervened node in a restricted SEM to conclude its invariance. In the next step, we will show how searching for such restricted SEMs for non-intervened nodes is simplified with this decomposition to equivalence classes.

\paragraph{d-separation property for invariance.} We establish the connection between d-separation in interventional graphs and the precision differences. Consider the augmented graph characterization of interventions presented in \cite{jaber2020causal}. A new node, $F$, is introduced to the graph to represent the interventional distribution. There are edges from $F$ to $i$ for any intervened node $i \in \mcI$ in the augmented graph. As there is no edge between $F$ and non-intervened node $j$, there exists a set $S$ that d-separates $F$ and $j$ in the augmented graph. This implies that the probability distribution of the node $j$ is invariant given $S \setminus \{j\}$, which in turn implies that both conditional mean and variance of the node $j$ does not change. Subsequently, $\sigma_{S,j}$ is invariant for this set $S$. Applying the results of \cite{pourahmadi2011covariance} and \cite{wang2018direct}, $[\Theta_S]_{j,j}= \sigma_{S,j}^{-2}$ is also invariant. Therefore, the set $S$ that d-separates $F$ and non-intervened $j$ results in $[\Delta_{\Theta_S}]_{j,j}=0$.

\emph{Processing equivalence classes.} We process equivalence classes $\mcA_1,\dots,\mcA_L$ individually, i.e., at stage $\ell$, we consider the nodes in $\mcA_\ell$. Let us define $\mcM_\ell = J_0 \cup \bigcup_{1\leq b < \ell}\mcA_b$. We will prove that for a non-intervened node $j \in \mcA_\ell$, we can determine its invariance via $2^{|\mcA_\ell|}$ PDE. Due to our ordering of the equivalence classes, any ancestor of $j$ in $S_\Delta$ will lie in either $\mcM_\ell$ or $\mcA_\ell$. Consider the set $S = \mcM_\ell \cup \An_\mcI(j) \cup \Pa(\An_\mcI(j))$ which is also of the form $\mcM_\ell \cup A$ for some $A \subseteq \mcA_\ell$.  Note that $S$ does not contain any descendant of $j$. 

We will use \emph{d-separation property for invariance} to show that this set $S$ yields $[\Delta_{\Theta_S}]_{j,j}=0$. Specifically, we will show that there does not exist a d-connecting path between the augmented node $F$ and $j$. Suppose the contrary and let $\pi: \langle F \rightarrow i \dots j \rangle$ be a d-connecting path where $i \in \mcI$. If $j$ has a tail end on $\pi$, there is a collider node $k$ on the path that is a descendant of $j$. Since $S$ does not contain any descendant of $j$, neither node $k$ nor its descendants are in $S$, and it blocks the path. Therefore, the path should be of the form $\langle F \rightarrow i \dots \rightarrow j \rangle$. If $i$ is a collider and not in $S$, it means it is not an ancestor of $S$. Therefore, its descendants are also not in $S$, and $i$ blocks the path. If $i$ is a collider and in $S$, it is either in $\mcM_\ell$ or in $\An_\mcI(j)$. In either case, the parent of $i$ on the path is also in $S$ and it blocks the path. If $i$ is not a collider, the path will be $\langle F \rightarrow i \rightarrow \dots \rightarrow j \rangle$. If $i$ is in $S$, it blocks the path. If $i$ is not in $S$, it is not an ancestor of $j$. Then, there is a collider $k$ on the path that is a descendant of $i$. Since $i$ is not in $S$, none of its descendants are neither in $S$. Therefore, $k$ blocks the path. We have ruled out all possible active paths and shown that there does not exist a d-connecting path between $F$ and $j$ for $S=\mcM_\ell \cup \An_\mcI(j) \cup \Pa(\An_\mcI(j))$. Subsequently, $[\Delta_{\Theta_S}]_{j,j}=0$ due to d-separation for invariance property. As we have noted before, set $S$ can be written as $S = \mcM_\ell \cup A$ for some $A \subseteq \mcA_\ell$, and we can check the existence of such $A$, i.e., whether $j$ is non-intervened by using PDE only $2^{|\mcA_\ell|}$ times. Formally, the \emph{process equivalence class} returns
\begin{align}
    \mcI_\ell = \{i: i \in \mcA_\ell \cap \mcI\}\ , \quad \mbox{and} \quad
    J_\ell &= \{j : j \in \mcA_\ell \cap \mcI^\C \}\ .
\end{align}
This concludes the proof that Algorithm~1 consistently estimates $\mcI$ set. \endproof

\begin{remark} \label{remark:pde_size}
 After forming $\mcA_1, \dots, \mcA_L$ classes with corresponding sets $J_0^{\mcA_1},\dots, J_0^{\mcA_L}$, consider a pair $\mcA_\ell,\mcA_\ell'$ where $1\leq \ell < \ell' \leq L$. Note that for any node pair $(u,v)$  where $u\in \mcA_\ell$ and $v\in \mcA_{\ell'}$, $u$ is not a descendant of $v$. Additionally, if $J_0^{\mcA_\ell} \not \subset J_0^{\mcA_{\ell'}}$, $u$ is not an ancestor of $v$. Hence, while considering $\mcA_\ell$ step of Algorithm~1, taking $\mcM_\ell = J_0^{\mcA_\ell} \cup \bigcup_{b \in \mcB_\ell}\mcA_b$ where  $\mcB_\ell \triangleq \{b: J_0^{\mcA_b} \subset J_0^{\mcA_\ell}, \;\; 1\leq b < \ell\}$ is equivalent to taking $\mcM_\ell = J_0 \cup \bigcup_{1\leq b < \ell}\mcA_b$. We use the former simplified approach to reduce the computational burden by having fewer nodes for subsequent $\Delta_\Theta$ estimates.
\end{remark}

\paragraph{{\bf Proof of Theorem \ref{theorem:parents}.}}
While processing a class $\mcA_\ell$ in Algorithm~\ref{alg:main_algorithm}, we declare a node $j$ non-intervened if there exist a set $A\subset \mcA_\ell$ such that $[\Delta_{\Theta_{\mcM_\ell \cup A}}]_{j,j}=0$. Note that there may exist more than one such $\mcM_\ell \cup A$, in which case we denote the smallest one by $\mcN_j$. 

Now, define $c_j \triangleq \ell$ for all $j \in \mcA_\ell$, where $\ell$ is the index of the equivalence class that contains node $j$. We have shown in Section \ref{sec:I-Markov} that finding $\{j \rightarrow i\}_{j \notin \mcI, i \in \mcI}$ is sufficient to update MEC into $\mcI$-MEC. Therefore, our goal for a non-intervened node is to find all of its intervened children. Consider $j \in J_{c_j}$ and $i \in \mcI_{c_i}$ such that $c_j \leq c_i$. If $i \in \mcN_j$, it immediately implies that $j$ is not a parent of $i$. Suppose that $i \notin \mcN_j$.

Consider $S = \mcM_{c_i} \cup \Pa(i) \cup \{i\}$ that is also of the form $\mcM_{c_i} \cup A$ for some $A \subseteq \mcA_{c_i}$. Therefore, we compute PDE for this $S$ in $c_i$-th stage of \emph{process equivalence class}. If $j \notin \Pa(i)$, all the paths $j \dots \rightarrow i$ are blocked with a parent of $i$ that is given in $S$. On the other hand, if the path ends with $\leftarrow i$, the path contains a collider node $k$ that is a descendant of $i$. Since $i$ is the youngest node in $S$, that collider $k$ blocks the path. Therefore, $[\Theta_S]_{j,i}=0$ and $[\Delta_{\Theta_S}]_{j,i}=0$ if $j \notin \Pa(i)$. From Assumption \ref{eq:assumption}, if $j \in \Pa(i)$, $[\Delta_{\Theta_S}]_{j,i}\neq 0$. Therefore, we identify all the non-intervened parents of intervened node $i$.

\paragraph{Orienting more edges.}
In addition to finding $\{j \rightarrow i\}_{j \notin \mcI, i \in \mcI}$, which is the main objective of Theorem \ref{theorem:parents}, we can also recover the edges $\{k \rightarrow i\}_{\{k,i\} \in \mcI, {c_{k}}\neq {c_{i}}}$. Consider nodes $k \in \mcI_{c_k}$ and $i \in \mcI_{c_i}$ such that $c_k < c_i$. In other words, $k$ and $i$ are both intervened but they belong to different equivalence classes. Similar to the previous case, by considering set $S = \mcM_{c_i} \cup \Pa(i) \cup i$, we obtain $[\Delta_{\Theta_S}]_{k,i}\neq 0$ if $k \in \Pa(i)$ and $[\Delta_{\Theta_S}]_{k,i}=0$ otherwise. Therefore, $k \notin \Pa(i)$, and we can orient all $k \rightarrow i$ edges if both nodes are intervened and belong to different equivalence classes.

\paragraph{{\bf Proof of Theorem \ref{theorem:sample_complexity}}.} We use the ADMM-based approach of \cite{jiang2018direct} as our PDE function to estimate $\Delta=\Thetao - \Thetat$. Theorem 1 of \cite{jiang2018direct} gives the sample complexity of this estimation as $O(M_\Sigma M_{\Gamma,\Gamma^T} d^4 \log p)$. In Theorem 3, we further assume that the product $M_\Sigma  M_{\Gamma,\Gamma^T} $ is bounded. Accordingly, with $n = O\left(\frac{d^4}{\varepsilon^2}\frac{\log p}{\delta}\right)$ samples, PDE's output $\hat\Delta$ satisfies $\|\hat\Delta-\Delta\|_{\infty} < \varepsilon$ with a probability at least $1-\delta$. We note that the conditions in Theorem \ref{theorem:sample_complexity} are given for the linear SEM over $[p]$ and the associated covariance matrices. If these conditions hold, they also hold for the restricted SEM over any $S \subset [p]$. Therefore, if we have $\|\hat \Delta_\Theta - \Delta_\Theta \|_\infty < \varepsilon$, we also have $\|\hat \Delta_{\Theta_S} - \Delta_{\Theta_S} \|_\infty < \varepsilon$ for any set $S$. Subsequently, we can threshold PDE outputs $\hat \Delta_{\Theta_S}$ by $\varepsilon$ to exactly recover the support of $\Delta_{\Theta_S}$ for any set $S$.

Note that Algorithm \ref{alg:main_algorithm} requires only the support of $\Delta_{\Theta_S}$ for a number of sets $S$. Therefore, with $n = O\left(\frac{d^4}{\varepsilon^2}\frac{\log p}{\delta}\right)$ samples, Algorithm \ref{alg:main_algorithm} identifies $\mcI$ with a probability at least $1-\delta$. We have shown in the proof of Theorem \ref{theorem:parents} that finding $\{j \rightarrow i\}_{j \notin \mcI, i \in \mcI}$ does not require any additional $\Delta_\Theta$ estimates. Therefore, with $n = O\left(\frac{d^4}{\varepsilon^2}\frac{\log p}{\delta}\right)$ samples, Algorithm \ref{alg:main_algorithm} also identifies the non-intervened parents of the intervened nodes $\{j \rightarrow i\}_{j \notin \mcI, i \in \mcI}$ with a probability at least $1-\delta$.
\endproof

\textcolor{black}{
We finally note that Corollary 1 of \cite{jiang2018direct} explicitly assumes that both $M_\Sigma$ and $M_{\Gamma,\Gamma^T}$ are bounded to remove $M_\Sigma M_{\Gamma,\Gamma^T}$ from the sample complexity. However, it can be readily relaxed to $M_\Sigma M_{\Gamma,\Gamma^T}<~+\infty$ since both terms always appear within the same product. We note that this relaxation brings about a significant level of flexibility in choosing covariance matrices. Indeed, this product is closely related to the condition number of the estimation problem. Two terms correspond to the norm of the inverse of Hessian of the optimization problem and the norm of the covariance, respectively. Product of these terms, the condition number, appears in similar matrix inference problems such as graphical lasso \cite{ravikumar2011high}. }

\section{Additional experiments}\label{sec:additional_experiments}

\subsection{Intervention recovery} \label{sec:additional_intervention_recovery}
We have compared the results of our algorithm and those of UT-IGSP for estimating intervention targets under shift intervention model in Section \ref{sec:intervention_recovery}. We expand the simulations to various settings in this subsection. Specifically, we report the results for shift intervention model with higher density $c=2.5$ in Table \ref{tab:shift_density25}, increased variance setting with $c=2.5$ in Table \ref{tab:increase_density25}, and randomized intervention setting with with $c=2.5$ in Table \ref{tab:perfect_density25}.

Our algorithm works well in all settings. Especially, increasing the dimension does not adversely affect accuracy and time complexity.

\begin{table}[h]
  \caption{$\mcI$ estimation in the shift intervention model - 50 repetitions with 5000 samples - density 2.5}
  \label{tab:shift_density25}
  \centering
  \begin{tabular}{lllllllll}
    \toprule
    \multicolumn{5}{c}{UT-IGSP (\cite{utigsp})} &
    \multicolumn{4}{c}{Algorithm \ref{alg:main_algorithm}}                   \\
    \cmidrule(r){1-9}
    p     & Precision & Recall & F1 & Time(s) & Precision     & Recall & F1 & Time(s) \\
    \midrule
    20 & 0.95  & 0.99 & 0.97 & 0.2 & 0.90 & 0.86 & 0.88 & 0.2     \\
    40 & 0.89  & 0.99 & 0.94 & 0.6 & 0.87 & 0.91 & 0.89 & 0.3    \\
    60 & 0.88  & 1 & 0.94 & 2.0 & 0.86 & 0.96 & 0.91 & 0.4     \\
    80 & 0.80  & 1 & 0.89 & 7.0 & 0.86 & 0.94 & 0.90 & 0.5     \\
    100 & 0.77 & 1 & 0.87 & 17.7 & 0.87 & 0.98 & 0.92 & 0.5  \\
    \bottomrule
  \end{tabular}
\end{table}
\begin{table}[h]
  \caption{$\mcI$ estimation in the increased variance model - 50 repetitions with 5000 samples - density 2.5}
  \label{tab:increase_density25}
  \centering
  \begin{tabular}{lllllllll}
    \toprule
    \multicolumn{5}{c}{UT-IGSP (\cite{utigsp})} &
    \multicolumn{4}{c}{Algorithm \ref{alg:main_algorithm}}                   \\
    \cmidrule(r){1-9}
    p     & Precision & Recall & F1 & Time(s) & Precision     & Recall & F1 & Time(s) \\
    \midrule
    20 & 0.90  & 0.99 & 0.95 & 0.2 & 0.89 & 0.86 & 0.87 & 0.2     \\
    40 & 0.85  & 1 & 0.92 & 0.6 & 0.87 & 0.93 & 0.90 & 0.3    \\
    60 & 0.88  & 1 & 0.93 & 2.4 & 0.89 & 0.97 & 0.92 & 0.3     \\
    80 & 0.80  & 1 & 0.89 & 5.8 & 0.86 & 0.97 & 0.91 & 0.4     \\
    \bottomrule
  \end{tabular}
\end{table}
\begin{table}[h]
  \caption{$\mcI$ estimation in the randomized intervention - 50 repetitions with 5000 samples - density 2.5}
  \label{tab:perfect_density25}
  \centering
  \begin{tabular}{lllllllll}
    \toprule
    \multicolumn{5}{c}{UT-IGSP (\cite{utigsp})} &
    \multicolumn{4}{c}{Algorithm \ref{alg:main_algorithm}}                   \\
    \cmidrule(r){1-9}
    p  & Precision & Recall & F1 & Time(s) & Precision & Recall & F1 & Time(s) \\
    \midrule
    20 & 0.92  & 1 & 0.96 & 0.2 & 0.86 & 0.91 & 0.88 & 0.2    \\
    40 & 0.82  & 1 & 0.90 & 0.7 & 0.88 & 0.94 & 0.91 & 0.3    \\
    60 & 0.81  & 1 & 0.90 & 2.8 & 0.84 & 0.96 & 0.90 & 0.5    \\
    80 & 0.74  & 1 & 0.85 & 8.4 & 0.86 & 0.92 & 0.89 & 0.6    \\
    \bottomrule
  \end{tabular}
\end{table}

\paragraph{Comparison with Ghoshal's algorithm \cite{ghoshal2019direct}.} 

Ghoshal's algorithm in \cite{ghoshal2019direct} is designed to estimate $\Delta_B$, and its performance critically hinges on the noise variances to be invariant. Even though it is not designed to return intervention targets, we can define the estimated intervention set of Ghoshal's algorithm as $\hat \mcI \triangleq \{ i : i, \; \exists \; j, (\Delta_B)_{j,i}\neq 0 \}$. We run our algorithm and Ghoshal's algorithm on the randomized intervention setting described in Section \ref{sec:intervention_recovery} and report the results in Table \ref{tab:perfect_ghoshal_comparison}. Expectedly, Ghoshal's algorithm does not perform well due to violation of the invariant noise variance assumption.
\begin{table}[h]
  \caption{$\mcI$ estimation in the randomized intervention model - 100 repetitions with 10000 samples - density 2.5}
  \label{tab:perfect_ghoshal_comparison}
  \centering
  \begin{tabular}{lllllllll}
    \toprule
    \multicolumn{5}{c}{Ghoshal \cite{ghoshal2019direct}} &
    \multicolumn{4}{c}{Algorithm \ref{alg:main_algorithm}}                   \\
    \cmidrule(r){1-9}
    p     & Precision & Recall & F1 & Time(s) & Precision     & Recall & F1 & Time(s) \\
    \midrule
    20 & 0.74  & 0.62 & 0.67 & <0.1 & 0.92 & 0.92 & 0.92 & 0.6    \\
    40 & 0.73  & 0.68 & 0.70 & 0.1 & 0.91 & 0.94 & 0.93 & 0.6    \\
    60 & 0.70  & 0.69 & 0.69 & 0.2 &  0.91 & 0.96 & 0.94 & 0.6     \\
    80 & 0.69  & 0.66 & 0.67 & 0.3 & 0.91 & 0.96 & 0.93 & 0.6     \\
    100 & 0.66 & 0.63 & 0.64 & 0.4 & 0.91 & 0.95 & 0.93 & 0.7 \\
    \bottomrule
  \end{tabular}
\end{table}

\paragraph{Increased number of samples.} Theorem \ref{theorem:consistency} states that our algorithm is consistent. Figure \ref{fig:ours_alone} shows that the performance of the algorithm increases significantly with the increased number of samples in all of the considered settings. We provide additional evidence of this fact. We generate 50 random graphs with density $c=2.5$ for each of the shift intervention, increased variance, and randomized intervention settings. We report the F1 scores for each setting with 5000, 10000, and 20000 samples in Table \ref{tab:more_samples}.

\begin{table}[h]
  \caption{$\mcI$ estimation with increased number of samples - 50 repetitions - density 2.5}
  \label{tab:more_samples}
  \centering
  \begin{tabular}{llllllllll}
    \toprule
    \multicolumn{1}{c}{} & 
    \multicolumn{3}{c}{Shift Intervention} &
    \multicolumn{3}{c}{Increased Variance} &
    \multicolumn{3}{c}{Randomized Intervention} \\
    \cmidrule(r){1-10}
    p & 5000 & 10000 & 20000 & 5000 & 10000 & 20000 & 5000 & 10000 & 20000 \\
    \cmidrule(r){1-10}
    40 & 0.87  & 0.90 & 0.91 & 0.95 & 0.96 & 0.96 & 0.90 & 0.92 & 0.94   \\
    60 & 0.90  & 0.92 & 0.93 & 0.93 &  0.96 & 0.97 & 0.91 & 0.93 & 0.95     \\
    80 & 0.90  & 0.91 & 0.94 & 0.93 & 0.97 & 0.98 & 0.91 & 0.94 & 0.95     \\
    100 & 0.93 & 0.94 & 0.96 & 0.94 & 0.97 & 0.97 & 0.89 & 0.93 & 0.92 \\
    \bottomrule
  \end{tabular}
\end{table}

\subsection{Causal structure learning}\label{sec:additional_causal_structure}
In Section \ref{sec:I-Markov}, we have shown that our method recovers the new information that can be gained through interventions. Hence, Algorithm \ref{alg:main_algorithm} refines the given MEC into the $\mcI$-MEC. Accordingly, we test our algorithm for the causal structure recovery task in this subsection. 

First, we take the correct CPDAG of $\mathcal{G}^{(1)}$ and apply our algorithm's findings to obtain $\mcI$-CPDAG. We run 100 realizations of Erd\H{o}s-R\'enyi graphs with $c=2$ and $10000$ samples. For different values of graph size $p$, we consider fixed target set size $|\mcI|=5$ or growing target set size $|\mcI|=p/10$. We report the results for recovery of $\mcI$-directed edges in Table \ref{tab:I_directed_edges_recovery}.

\begin{table}[h]
  \caption{Recovery of $\mcI$-directed edges in the increased variance model}
  \label{tab:I_directed_edges_recovery}
  \centering
  \begin{tabular}{lllllllll}
    \toprule
    \multicolumn{5}{c}{$|\mcI|=5$} &
    \multicolumn{4}{c}{$|\mcI|=p/10$}                   \\
    \cmidrule(r){1-9}
    p  & Precision & Recall & F1 & Time(s) & Precision & Recall & F1 & Time(s) \\
    \midrule
    40 & 0.69  & 0.93 & 0.80  & 0.15  & 0.73  & 0.94 & 0.82  & 0.11  \\
    60 & 0.73 & 0.93 & 0.82 & 0.24 & 0.73 & 0.93 & 0.82 & 0.25\\
    80 & 0.75 & 0.93 & 0.83 & 0.28 & 0.73 & 0.96 & 0.83 & 0.45  \\
    100 & 0.82 & 0.97 & 0.89 & 0.42  & 0.72 & 0.93 & 0.81 & 0.86  \\
    \bottomrule
  \end{tabular}
\end{table}

Next, we consider recovering the non-intervened parents of the intervened nodes, i.e., $\{j \rightarrow i\}_{j \notin \mcI, i \in \mcI}$. We note that we do not use any given MEC information in this setting. Therefore, a comparison with UT-IGSP algorithm becomes feasible. We report the results for $|\mcI|=5$ in Table \ref{tab:setting_b_1}. Similar to the intervention recovery task, our algorithm's runtime does not suffer from increasing the dimension while the runtime of UT-IGSP grows very quickly.

\begin{table}[h]
  \caption{Recovery of non-intervened parents of intervened nodes}
  \label{tab:setting_b_1}
  \centering
  \begin{tabular}{lllllllll}
    \toprule
    \multicolumn{5}{c}{UT-IGSP (\cite{utigsp})} &
    \multicolumn{4}{c}{Algorithm \ref{alg:main_algorithm}}                   \\
    \cmidrule(r){1-9}
    p  & Precision & Recall & F1 & Time(s) & Precision & Recall & F1 & Time(s) \\
    \midrule
    20 & 0.76  & 0.98 & 0.86 & 0.32 & 0.81 & 0.81 & 0.81 & 0.15    \\
    40 & 0.82  & 0.98 & 0.89 & 2.30 & 0.85 & 0.79 & 0.82 & 0.22   \\
    60 & 0.84  & 0.98 & 0.91 & 10.11  & 0.88 & 0.85 & 0.86 & 0.26    \\
    80 & 0.89  & 0.99 & 0.93 & 32.97 & 0.92 & 0.78 & 0.85 & 0.28    \\
    \bottomrule
  \end{tabular}
\end{table}

\subsection{Application to real data}\label{sec:additional_real_data}
\begin{figure}[h]
    \centering
    \begin{subfigure}{0.45\linewidth}
        \centering
        \includegraphics[width=1.0\linewidth]{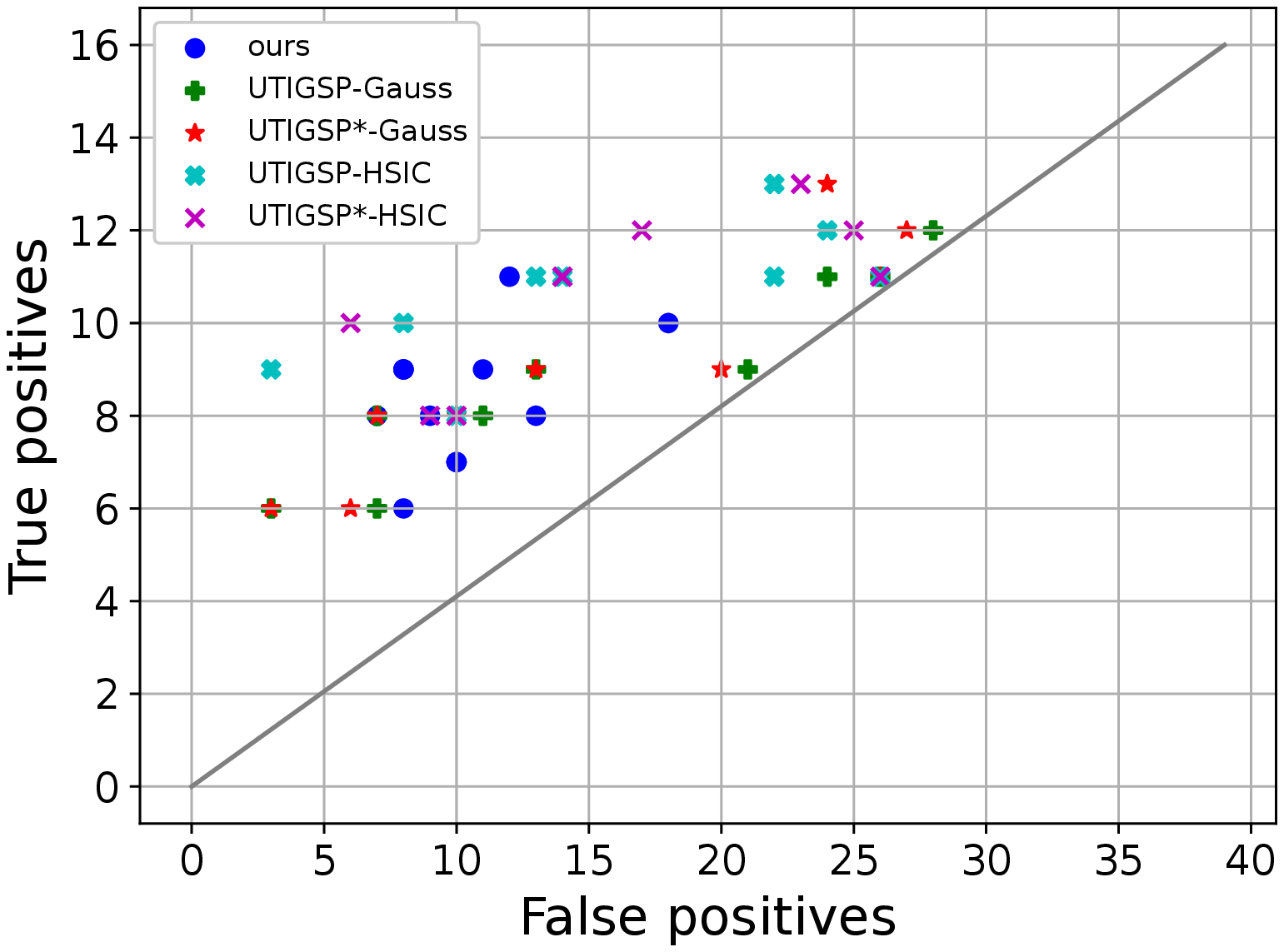}
        \caption{Sachs \cite{SACHSKaren2005Cpnd} dataset skeleton recovery}
        \label{fig:sachs_skeleton}
    \end{subfigure}
    \begin{subfigure}{0.45\linewidth}
        \centering
        \includegraphics[width=1.0\linewidth]{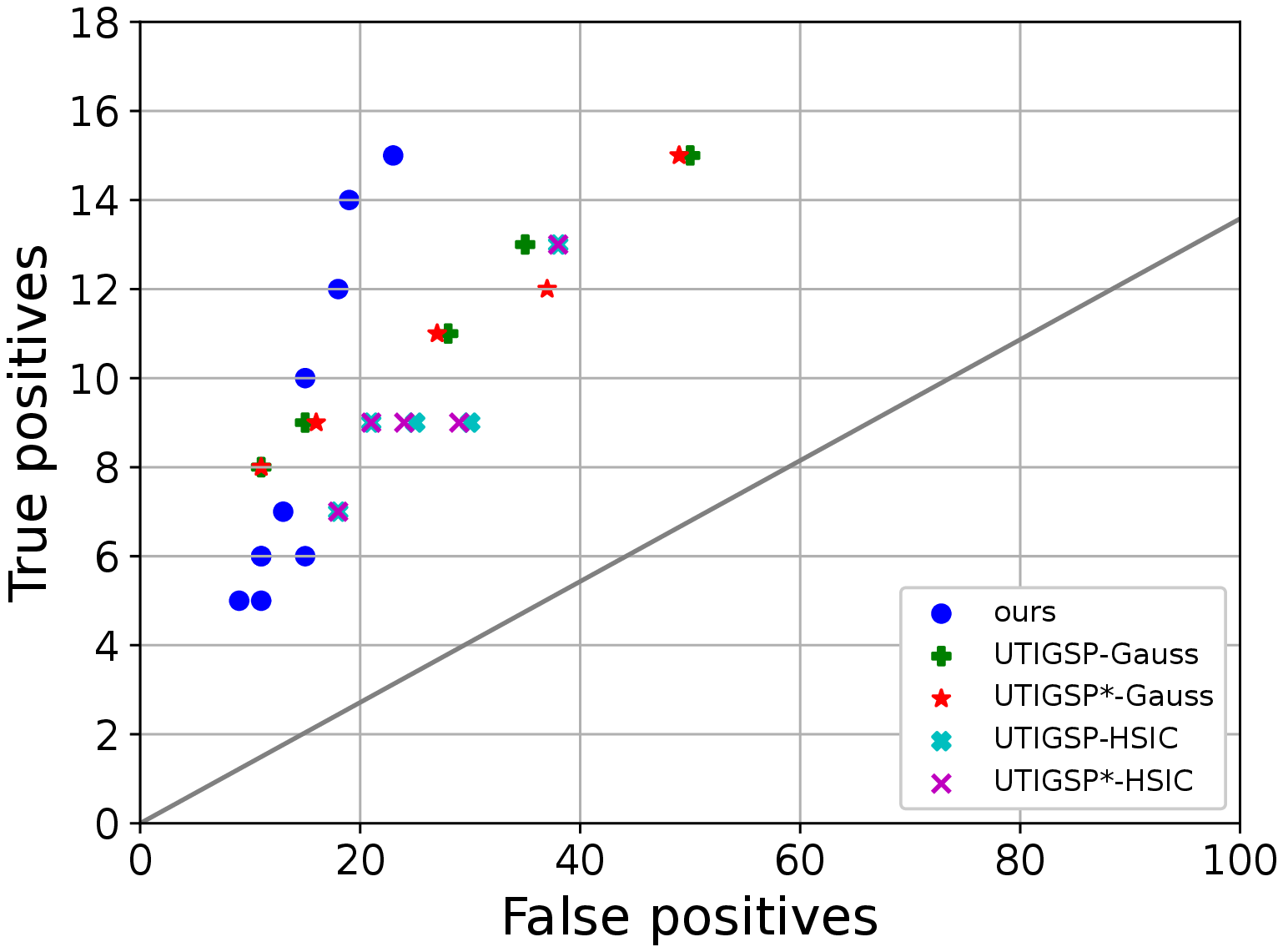}
        \caption{Dixit \cite{perturb-seq} dataset skeleton recovery}
        \label{fig:dixit_skeleton}
    \end{subfigure}    
    \caption{ROC curves for skeleton recovery. The solid grey line corresponds to random guessing.}
    \label{fig:real_skeleton}
\end{figure}

We have investigated directed edge recovery results for two real biological datasets in Section \ref{sec:real_data}. In this subsection, we give the skeleton recovery results for the same datasets. Figure \ref{fig:real_skeleton} illustrates that our observations from the directed edge recovery hold for the skeleton recovery as well. Comparison of figures \ref{fig:real_directed} and \ref{fig:real_skeleton} reveals that our algorithm orients fewer number of edges incorrectly with respect to UT-IGSP algorithm.

\noindent\textbf{Hyperparameters.} We have defined the regularization parameters $\lambda_1,\lambda_2$, and $\lambda_3$ for our algorithm and cut-off value $\alpha$ for UT-IGSP in Section \ref{sec:simulations}. Specifically, we have used $\lambda_1 \in [0.1,0.3]$, $\lambda_2=0.2$, and $\lambda_3 \in [0.05,0.2]$ for Algorithm \ref{alg:main_algorithm}, and $\alpha \in [0.0001,0.5]$ for UT-IGSP while creating figures \ref{fig:sachs_directed} and \ref{fig:sachs_skeleton}. Similarly, we have used $\lambda_1=0.1$, $\lambda_2=0.05$, and $\lambda_3 \in [0.005,0.1]$ for Algorithm \ref{alg:main_algorithm}, and $\alpha \in [0.005,0.1]$ for UT-IGSP while creating figures \ref{fig:dixit_directed} and \ref{fig:dixit_skeleton}.

\subsection{Computational complexity}\label{sec:additional_computational_complexity}
We have stated in Section \ref{sec:algorithm} that the computational complexity of our algorithm is exponential in the size of the largest equivalence class, ${\max |\mcA_\ell|}$. This can be as large as $p_\Delta$ in some extreme examples. One possible scenario for this case is if the parents of intervention targets are also intervened. In this case, $J_0$ will be the empty set and all nodes in $S_\Delta$ will belong to the same group. However, this requires the interventions to concentrate in one neighborhood such that parents of the intervened nodes will also be intervened. In reality, such scenarios happen rarely, and interventions are generally distributed. 

\begin{figure}[t]
    \centering
    \begin{subfigure}{0.45\linewidth}
        \centering
        \includegraphics[width=1\linewidth]{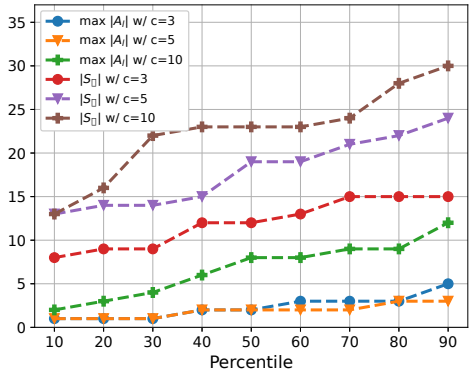}
        \caption{$p=100, |\mcI|=5$}
        \label{fig:p100_I5_sizes}
    \end{subfigure}
    \begin{subfigure}{0.45\linewidth}
        \centering
        \includegraphics[width=1\linewidth]{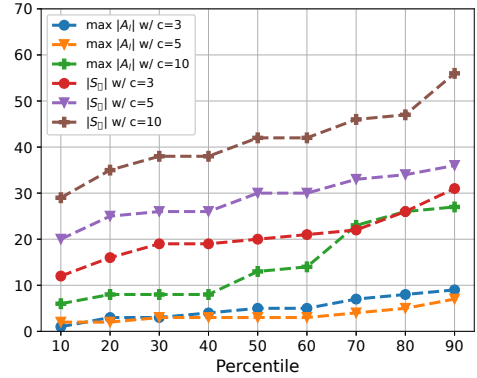}
        \caption{$p=100, |\mcI|=10$}
        \label{fig:p100_I10_sizes}
    \end{subfigure}    
    \caption{Exponential factor in the computational complexity of our algorithm, $\max_\ell |\mcA_\ell|$, is much smaller than the size of the affected nodes $p_\Delta=|S_\Delta|$. x-axis shows the percentile values over 1000 different random DAG instances. Largest class size $\max_l |\mcA_\ell|$ and $p_\Delta$ are plotted for three different density values.}
    \label{fig:p100_sizes}
\end{figure}

We generate 1000 instances of random graphs with $p=100$, various densities, and target set sizes to demonstrate the much smaller size of $\mcA_\ell$ groups with respect to $S_\Delta$. Figure \ref{fig:p100_sizes} illustrates that $\max |\mcA_\ell|$ is much smaller than $p_\Delta$. Indeed, Fig. \ref{fig:p100_sizes} also shows the limitations of some of the related work that has computational complexity exponential in $p_\Delta$ strictly. For instance, for $p=100,|\mcI|=5$, and $c=5$ in Fig. \ref{fig:p100_I5_sizes}, the $90\%$-th percentile of $p_\Delta$ is $25$, whereas $\max_\ell |\mcA_\ell|$ is only $4$. Gains of our algorithm become more dramatic when the target set is larger. For instance, for $p=100, |\mcI|=10$, and $c=5$ in Fig. \ref{fig:p100_I10_sizes}, the $50\%$-th percentile of $p_\Delta$ is $34$, whereas the $90\%$-th percentile of $\max_\ell |\mcA_\ell|$ is only~$10$. Therefore, our algorithm can scale up to higher dimensions.

We finally comment on the computational complexity of the PDE routine. The ADMM-based PDE algorithm of \cite{jiang2018direct} has $O(p^3)$ complexity. We note that we run PDE with all $[p]$ nodes only once during the $S_\Delta$ estimation in Step 1. Hence, the estimation with $O(p^3)$ complexity will only be performed once. The rest of the PDE instances require much smaller number of nodes as stated in Remark \ref{remark:pde_size}. We note that a related study in \cite{ghoshal2019direct} uses another PDE algorithm that has complexity $O(p^4)$. Reducing it to $O(p^3)$ is a significant gain, which allows us to process hundreds of nodes. 

\end{document}